\documentstyle{ar}
\input epsf
\begin{document}


\title{THE QUANTUM PHYSICS OF BLACK HOLES: Results from String Theory
\footnote{With permission from the Annual Review of Nuclear and Particle Science.
Final version of this material appears in the Annual Review of Nuclear and
Particle Science Vol. 50, published in December 2000 by Annual Reviews,
http://AnnualReviews.org.}
}  
\markboth{Das \& Mathur}{D-Branes and Black Holes}

\author{Sumit R. Das\affiliation{Tata Institute of Fundamental Research,
Homi Bhabha Road, Mumbai 400005, India; e-mail: das@theory.tifr.res.in}
Samir D. Mathur\affiliation{Department of Physics,
Ohio State University, Columbus, Ohio 43210; e-mail: mathur@mps.ohio-state.edu}}

\begin{keywords}
black holes, information loss, string theory, D-branes, holography
\end{keywords}

\begin{abstract}
We review recent progress in our understanding  of the physics of
black holes.
In particular, we discuss the ideas from string theory that explain
the entropy of black holes from a counting of microstates of the hole,
and the related derivation of unitary Hawking radiation from such holes.
\end{abstract}

\maketitle

\def\ben{\begin{equation}}
\def\een{\end{equation}}
\def\bea{\begin{eqnarray}}
\def\eea{\end{eqnarray}}
\def\nn{\nonumber}
\def\bpsi{{\bar \psi}}
\def\bsigma{{\vec{\sigma}}}
\def\talpha{{\tilde{\alpha}}}
\def\tb{{\tilde{b}}}
\def\td{{\tilde{d}}}
\def\ba{{\bar{a}}}
\def\g{{\gamma}}
\def\LD{{L^{(D)}}}
\def\LS{{L^{(S)}}}
\def\TD{{T^{(D)}}}
\def\TS{{T^{(S)}}}
\def\half{{1\over 2}}
\def\noi{\noindent}

\section{INTRODUCTION}

Black holes present us with a very deep paradox. The path to
resolving this paradox may well be the path to
a consistent unified theory of matter and quantized gravity.

In classical gravity, a black hole is a classical solution of the
equations of motion such that there is a region of spacetime that is causally
disconnected from asymptotic infinity (see e.g.\ Reference~\cite{books}).
The boundary of such a region
is called the event horizon.

Consider a large collection of low-density matter, in an asymptotically flat spacetime. For simplicity, we take the starting
configuration to be spherically symmetric and nonrotating (these
restrictions do not affect the nature of the paradox that
emerges). This ball of matter will collapse toward smaller radii under
its self-gravitation. At some point, the matter will pass through a
critical radius, the Schwarzschild radius $R_s$, after which its
further collapse cannot be halted, whatever the equation of
state. The final result, in classical general relativity, is that the
matter ends up in an infinite-density singular point, while the metric
settles down to the Schwarzschild form
\begin{equation}
ds^2=-(1-{2G_NM\over rc^2})dt^2+(1-{2G_N M\over rc^2})^{-1}dr^2+r^2d\Omega^2.
\label{eq:one}
\end{equation}
Here $G_N$ is  Newton's constant of gravity, and $c$ is the speed of light.
The horizon radius of this hole is
\begin{equation}
R_s={2G_NM\over c^2}\rightarrow {2M},
\end{equation}
where the last expression arises after we set $G_N=1$, $c=1$.
(In what follows, we adopt these units unless otherwise explicitly indicated;
we also set $\hbar=1$.)

Classically, nothing can emerge from inside the horizon to the
outside.  A test mass $m$ has effective energy zero if it is placed at
the horizon; it has rest energy $mc^2$, but a negative
gravitational potential energy exactly balances
this positive
contribution. For a rough estimate  of the horizon size, we may put
this negative energy to
be the Newtonian value
$-G_NMm/r$, for which $ R_s \sim G_NM/c^2$.

It may appear from the above that the gravitational fields at the
horizon of a black hole are very large. This is not true. For a
neutral black hole of mass $M$, the magnitude of the curvature
invariants, which are the measure of local gravitational forces, is
given by
\ben
| {\cal R}| \sim {G_N M \over r^3}.
\een
Thus, at the horizon $r = r_H = 2G_N M$, the curvature scales as
$1/M^2$. As a result, for black holes with masses $M \gg G_N^{-1/2}$,
the curvatures are very small and the spacetime is locally rather close
to flat spacetime. In fact, an object falling into a black
hole will not experience any strong force as it crosses the horizon.
However, an asymptotic observer watching this object will see that it
takes an infinite time to reach the horizon. This is because there is an infinite gravitational red-shift between the horizon
and the asymptotic region.

An important point about black hole formation is that one does
not need to crush matter to high densities to form a black hole. In
fact, if the hole has mass $M$, the order of magnitude of the density
required of the matter is
\begin{equation} \rho\sim {M\over
     R_s^3}\sim {1\over M^2}.
\end{equation}
Thus, a black hole of the  kind believed to exist at the center of our
galaxy ($10^8$ solar masses) could form from a ball with the density
of water. In fact, given any density we choose, we can make a
black hole if we take a sufficient total mass with that density. This
fact makes it very hard to imagine a theory in which black holes do not
form at all because of some feature of the interaction between the
matter particles. As a consequence, if black holes lead to a paradox,
it is hard to bypass the paradox by doing away with black holes
in the theory.

It is now fairly widely believed that black holes exist in nature.
Solar-mass  black holes can be endpoints of stellar
evolution, and supermassive black holes ($\sim 10^5 -10^9$ solar
masses) probably exist at the centers of galaxies.  In some
situations, these holes accrete matter from
their surroundings, and
the collisions among these infalling particles
create very powerful sources of radiation
that are believed to be the source of the high-energy output of
quasars. In this article, however, we are not concerned with any
of these astrophysical issues. We concentrate instead on the
quantum properties of isolated black holes, with a view toward
understanding the problems that arise as issues of principle when
quantum mechanical ideas are put in the context of black holes.

For example, the Hawking radiation process discussed below
is a quantum process that is much weaker than the radiation from the
infalling matter mentioned above, and it would be almost impossible to
measure even by itself.
(The one possible exception is the Hawking
radiation at the last stage of quantum evaporation. This radiation
emerges in a sharp burst with a universal profile, and there are
experiments under way to look for such radiation from very small
primordial black holes.)

\subsection{The Entropy Problem}

Already, at this stage, one finds what may be called the entropy
problem.  One of the most time-honored laws in physics has been the
second law of thermodynamics, which states that the entropy of matter
in the Universe cannot decrease.
But with a black hole present in the
Universe, one can imagine the following process.
A box containing some gas, which has a certain entropy, is dropped into a large black hole. The metric of the black hole then soon
settles down to the Schwarzschild form above, though with a larger
value for $M$, the black hole mass. The entropy of the gas has
vanished from view, so that if we only count the entropy that we can
explicitly see, then the second law of thermodynamics has been
violated!

This violation of the second law can be avoided if one associates an
entropy to the black hole itself. Starting with the work of Bekenstein
\cite{bek}, we now know that if we associate an entropy
\begin{equation}
S_{{\mathrm BH}}={A_H\over 4G_N}
\label{eq:two}
\end{equation}
with the black hole of horizon area $A_H$,
then in any Gedanken experiment in which we try to lose entropy
down the hole, the increase in the black hole's attributed entropy is
such that
$${d\over dt}(S_{{\rm matter}}+S_{{\mathrm BH}})\ge 0$$
(for an analysis of such Gedanken experiments, see e.g.\ \cite{wald}).
Furthermore, an ``area theorem''  in general relativity
states that in any classical process, the total area of all black
holes cannot decrease. This statement is rather reminiscent of
the statement of the second law of thermodynamics---the entropy of
the entire Universe can  never decrease.

Thus the proposal (Equation \ref{eq:two}) would appear to be a nice one,
but now we encounter the following problem.  We would also like to believe
on general grounds
that thermodynamics can be understood in terms of
statistical mechanics; in particular,
the entropy $S$ of any system is given by
\ben
S = {\rm log}~ \Omega,
\label{eq:none}
\een
where $\Omega$ denotes the number of states of the system for a
given value of the macroscopic parameters. For a black hole of one
solar mass, this implies that there should be $10^{10^{78}}$ states!

But the metric (Equation \ref{eq:one}) of the hole suggests a unique state for the
geometry of the configuration. If one tries to consider small
fluctuations around this metric, or adds in, say, a scalar field in
the vicinity of the horizon, then the extra fields soon flow off to
infinity or fall into the hole, and the metric again settles down to
the form of Equation~\ref{eq:one}.

If the black hole has a unique state, then the entropy should be $\ln
1=0$, which is not what we expected from Equation~\ref{eq:two}. The idea that
the black hole configuration is uniquely determined by its mass (and
any other conserved charges) arose from studies of many simple examples of the matter fields.
This idea of uniqueness was encoded in the
statement ``black holes have no hair.'' (This statement is not strictly
true when more general matter fields are considered.)
It is a very interesting and
precise requirement on the theory of quantum gravity plus matter that
there be indeed just the number   (Equation~\ref{eq:two}) of microstates
corresponding to a given classical geometry of a black hole.

\subsection{Hawking Radiation}

If black holes have an entropy $S_{{\mathrm BH}}$ and an energy equal to the
mass $M$, then if thermodynamics were to be valid, we would expect
them to have a temperature given by
\begin{equation}
TdS=dE=dM. ~~~ \label{bthree}
\end{equation}
For a neutral black hole in four spacetime dimensions, $A_H = 4\pi (2G_NM)^2$,
which gives
\ben
T=({dS\over dM})^{-1}={1\over 8\pi G_N M}.
\label{eq:ntwo}
\een
Again assuming thermodynamical behavior, the above statement implies
that if the hole can absorb photons at a given wave number $k$ with absorption
cross section $\sigma (k) $, then it must also radiate at the same
wave number at the rate
\begin{equation}
\Gamma (k) = {\sigma (k) \over \displaystyle{e^{\hbar |k|\over kT}-1}}~
{d^dk\over (2\pi)^d}.
\label{eq:four}
\end{equation}
In other words, the emission rate is given by the absorption cross
section 
multiplied by
a   standard
thermal factor (this factor would have a plus sign in place of the minus
sign if we were considering fermion emission) and a  phase
space factor that counts the number of  states in the wave number
range ${\vec k}$ and ${\vec k} + d{\vec k}$. ($d$ denotes the
number of spatial dimensions.)

Classically, nothing can come out of the black hole horizon, so it is tempting
to say that no such radiation is possible.
However, in 1974, Hawking \cite{hawking}
found that if the quantum behavior
of matter fields is considered, such radiation is possible.
The vacuum for the matter fields has
fluctuations, so that pairs of particles and antiparticles are
produced and annihilated continuously.
In normal spacetimes, the pair annihilates quickly in a time set
by the uncertainty principle. However, in a black hole background,
one member of this pair can fall into the hole,
where it has a net negative energy, while the other member of the pair
can escape to infinity as real positive energy radiation
\cite{hawking}. The profile of this radiation is found to be thermal,
with a temperature given by Equation~\ref{eq:ntwo}.

Although we have so far discussed the simplest black holes, there are
black hole solutions that carry charge and angular momentum. We can
also consider generalizations of general relativity to arbitrary
numbers of spacetime dimensions (as will be required below) and
further consider other matter fields in the theory.

It is remarkable that the above discussed thermodynamic properties of
black holes
seem to be universal.
The leading term in the  entropy is in fact
given by Equation~\ref{eq:two} for all black holes of all kinds in any number of
dimensions. Furthermore, the temperature is given in terms of
another geometric quantity called the surface gravity at the horizon,
$\kappa$,
which is the acceleration
felt by a static object at the horizon as measured from the asymptotic region. The precise relation---also universal---is
\ben
T = {\kappa \over 2\pi}.
\label{eq:nthree}
\een

\subsection{The Information Problem}

``Hawking radiation'' is produced from the quantum fluctuations of the
matter vacuum, in the presence of the gravitational field of the
hole. For black holes of masses much larger than the scale set by
Newton's constant, the gravitational field near the horizon, where the
particle pairs are produced in this simple picture, is given quite
accurately by the classical metric of Equation~\ref{eq:one}.  The curvature
invariants at the horizon are all very small compared with the Planck
scale, so quantum gravity seems not to be required. Further, the
calculation is insensitive to the precise details of the matter that
went to make up the hole.
Thus, if the hole completely evaporates away,
the final radiation state cannot have any significant information
about the initial matter state. This circumstance would contradict the
assumption in usual quantum mechanics
that the final state of any time
evolution is related in a one-to-one and onto fashion to the initial
state,
through a unitary evolution operator. Worse, the final state is
in fact not even a normal quantum state. The outgoing member of a pair
of particles created by the quantum fluctuation is in a mixed state
with the member that falls into the hole, so that the outgoing
radiation is highly ``entangled'' with whatever is left behind at the
hole. If the hole completely evaporates away, then this final state is
entangled with ``nothing,'' and we find that the resulting system is
described not by a pure quantum state but by a mixed state.

If the above reasoning and computations are correct, one confronts a set
of alternatives, none of which are very palatable
(for a survey see e.g.\ Reference~\cite{gid}). The
semiclassical reasoning used in the derivation of Hawking radiation
cannot say whether the hole continues to evaporate after it reaches
Planck size, since at this point quantum gravity
would presumably have to be important. The hole may not completely
evaporate away but leave a ``remnant''
of Planck size. The radiation sent off to
infinity will remain entangled with this remnant. But this entanglement entropy
is somewhat larger \cite{zurek} than the black hole entropy $S_{{\mathrm BH}}$,
which is a very large
number (as we have seen above). Thus, the remnant will have to have a very large
number of possible states, and this number will grow to infinity as the mass of
the initial hole is taken to infinity. It is uncomfortable to have
a theory in
which a particle of bounded mass (Planck mass) can have an
infinite number of configurations. One might worry that in any quantum process,
one can have loops of this remnant particle,
and this contribution will diverge,
since the number of states of the remnant is infinite. But it has been argued
that remnants from holes of increasingly large mass might couple to any given
process with correspondingly smaller strength, and then such a
divergence can be avoided.

Another possibility, advocated most strongly by Hawking, is that the hole does
evaporate away to nothing, and the passage from an intial pure state to a final
mixed state is a natural process in any theory of quantum gravity. In this view,
the natural description of states is in fact in terms of density matrices, and
the pure states of quantum mechanics that we are used to thinking
about are only
a special case of this more general kind of state. Some investigations of this
possibility have suggested, however, that giving up the purity of
quantum states
causes difficulties with maintaining energy conservation in virtual
processes (\cite{banks}; for a counterargument, see Reference~\cite{flamingo}).

The possibility that would best fit our experience of physics in
general
would be that the Hawking radiation does manage to
carry out the information of the collapsing matter \cite{thooft}. The
hole could then
completely evaporate away, and yet the process would be in line with the
unitarity of quantum mechanics. The Hawking radiation from the black
hole
would not fundamentally differ from
the radiation from a
lump of burning coal---the information of the atomic structure of the
coal is contained, though it is difficult to decipher,
in the radiation
and other products that emerge when the coal burns away.

\subsection{Difficulties with Obtaining Unitarity}

Let us review briefly the difficulties with having the radiation carry out the
information.

To study the evolution, we  choose a foliation of the spacetime by
smooth spacelike hypersurfaces.
This requires that the spatial slices be smooth and that the embedding of
neighboring slices changes in a way that is not too sharp.

As we evolve along this foliation,
we see the matter fall in toward
the center of the hole, while we see the radiation collect at spatial
infinity. It is important to realize that the information in the
collapsing matter cannot also be copied into the radiation---in
other words, there can be no quantum ``Xeroxing.'' The reason is as follows.
Suppose the evolution process makes two copies of a state
$$|\psi_I \rangle \rightarrow |\psi_I \rangle \times|\psi_i \rangle,$$
where the $|\psi_i \rangle$ are a set of basis states.
Then, as long as the linearity of quantum mechanics holds, we will find
$$|\psi_1 \rangle +|\psi_2 \rangle ~\rightarrow
|\psi_1 \rangle \times|\psi_1 \rangle +|\psi_2 \rangle \times|\psi_2 \rangle $$
and not
$$|\psi_1 \rangle +|\psi_2 \rangle ~\rightarrow  (|\psi_1 \rangle +|\psi_2 \rangle )\times(|\psi_1 \rangle +|\psi_2 \rangle ).$$
Thus, a general state cannot be ``duplicated'' by any quantum process.

\begin{figure}
\epsfysize=6cm \epsfbox{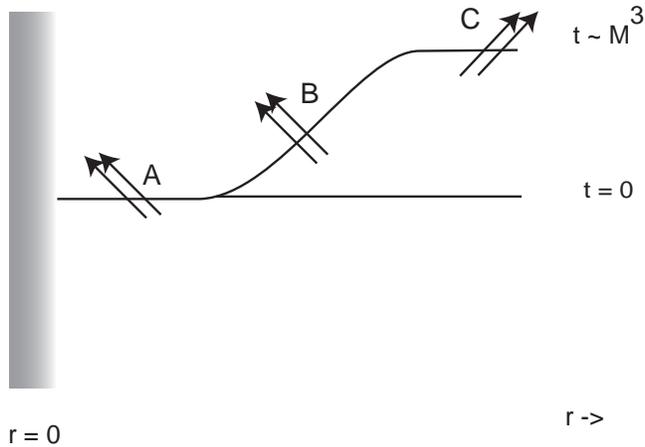}
\caption{Foliation of the black hole spacetime.}
\end{figure}

Figure 1 shows the spacetime in a symbolic way.  We use a
foliation of spacetime by the following kind of spacelike
hypersurfaces.  Away from the black hole, say for $r>4M$, we let the
hypersurface be a $t=t_0$ surface (this description uses the Schwarzschild
coordinates of Equation~\ref{eq:one}).  Inside the black hole, an
$r=$constant
surface is spacelike; let us choose $r=M$ so that this
part of the surface is neither close to the horizon ($r=2M)$ nor close to
the singularity ($r=0$).  This part of the hypersurface will extend
from some time $t=0$
near the formation of the hole to the value
$t=t_0$.  Finally, we can connect these two parts of the hypersurface
by a smooth interpolating region that is spacelike as well.  Each of
the spacelike hypersurfaces shown in

Figure~1
is assumed to be of
this form. The lower one has $t=0$, whereas the upper one corresponds to
a time $t_0\sim M^3$, where a mass $\sim M$ has been evaporated away as
radiation. We assume, however, that at $t_0$ the black hole is nowhere
near its endpoint of evaporation, either by assuming that a slow dose
of matter was continually fed into the black hole to maintain its size
(the simplest assumption) or by considering a time $t_0$ where say a
quarter of the hole has evaporated (and modifying the metric to
reflect the slow decrease of black hole mass).

On the lower hypersurface, we have
on the left
the matter that fell in
to make the hole. There is no radiation yet, so there is nothing else
on this hypersurface.  Let us call this matter ``A.'' On the upper
hypersurface, we expect the following sources of stress energy, in a
semiclassical analysis of the Hawking process. On the left, we will
still have the matter that fell in to make the hole, since this part
of the surface is
common to both hypersurfaces. On
the extreme right, we will have the Hawking radiation that has emerged
in the evaporation process; let us call this ``C.''  In the middle
are the infalling members of the particle-antiparticle pairs. These
contribute  a negative value to the total mass of the system because
of the way the
hypersurface is oriented with respect to the coordinate $t$---this
maintains overall energy conservation in the process. Let us call
this part of the state ``B.''

The semiclassical process gives a state for the light matter fields,
which is entangled between components B and C. On the other hand,
components A and B are expected to somehow vanish together
(or leave a Planck mass remnant), since their energies cancel each
other. At the end of the process, the radiation C will have the energy
initially present in A. But since C will be entangled with B, the
final state will not be a pure state of radiation.

We can now see explicitly the difficulties with obtaining in any
theory a unitary description of the process of black hole formation and
evaporation. In a general curved spacetime, we should be able
to evolve our hypersurfaces by different amounts at different
points---this is the ``many-fingered time'' evolution of general relativity
extended to include the quantum matter on the spacetime.
By using an appropriate choice of this evolution,
we have captured both
the infalling matter A and the outgoing radiation C on the same
spacelike hypersurface. If we want the radiation C to carry the
information of the matter A, then we will need ``quantum xeroxing,''
which, as mentioned above, cannot happen if we accept the
principle of superposition of quantum mechanics. It would have been
very satisfactory
if we just could not draw
a smooth hypersurface like
the upper one in Figure~1, a hypersurface that includes both the infalling
matter and the outgoing radiation. For example, we could have hoped
that any such surface would need to be non-spacelike at some point, or
that it would need a sharp kink in one or more places.  But it is easy to see
from the construction of surfaces described above that all the
hypersurfaces in the evolution are smooth. In fact, the later one is in
some sense just a time translate
of the earlier one---the part $t={\rm
constant}$ in each surface has the same intrinsic (and extrinsic)
geometry for each hypersurface, and the segment that connects this
part to the $r={\rm constant}$ part can be taken to be the same as
well.
The only difference between the hypersurfaces is that the later
one has a larger $r={\rm constant}$ part. One can further check that the
infalling matter has finite energy along each hypersurface and that scalar
quantities such as $dr/ds$ are bounded and smooth along each surface ($s$
is the proper length along the surface). In the above calculations,
spacetime was treated classically, but the conclusions do not change
even if we let the evolution of spacelike surfaces be described by
the Wheeler--de~Witt equation,
which gives a naive quantization of gravity;
quantum fluctuations of the spacetime may appear large in certain
coordinates \cite{klmo}, but  such effects cancel out in the
computation of Hawking radiation
\cite{km}.

It thus appears that in order to have unitarity one needs a nonlocal mechanism
(which operates over macroscopic distances $\sim M$) that moves the
information from A to C. Even though the spacetime appears to have no
regions of Planck-scale curvature, we must alter our understanding of
how information in one set of low-energy modes (A) moves into another
set of low energy modes (C). A key point appears to be that,
in the semiclassical calculation, the radiation C emerges from modes of
the quantum field that in the past had a frequency much higher than
Planck frequency. A naive model of the quantum field would have
these
modes at all frequencies, but if the complete theory of matter and
gravity has an inbuilt
cutoff at the Planck scale, then the radiation C
must have had its origins somewhere else---possibly in some nonlocal
combination of modes with sub-Planckian energy. If some such
possibility is true, we would obtain unitarity, while also obtaining
some nontrivial insight into the high-energy structure of the quantum
vacuum.

Basic to such an approach would be some way of understanding a black
hole as a complicated version of usual matter states, and not as an
esoteric new object that must be added to a theory of ``regular''
matter.
It would still be true that the final state of a system changes character significantly when its density changes from that of a star, for instance, to the density at which it collapses to form a black hole,
but the resulting hole should still be described by the
same essential principles of quantum mechanics, density of states,
statistical mechanics, etc, as any other matter system. As we show
below, string theory provides not only a consistent theory of
quantized gravity and matter, but also a way
of thinking about black holes as quantum states of the matter variables
in the theory.

\section{STRING THEORY AND SUPERGRAVITY}

In a certain regime of parameters, string theory
is best thought of
as a theory of interacting elementary strings (for expositions of superstring theory, see Reference~\cite{strings}). The basic scale is set by
the string tension $T_s$ or, equivalently, the ``string length''
\ben
l_s = {1\over {\sqrt{2\pi T_s}}}.
\label{eq:szero}
\een
The quantized harmonics of a string represent particles of various
masses and spins, and the masses are typically integral multiples
of $1/l_s$. Thus, at energies much smaller than $1/l_s$, only the
lowest harmonics are relevant. The interaction between strings is
controlled by a dimensionless
string coupling $g_s$, and the above description of the
theory in terms of propagating and interacting strings is a good
description
when $g_s \ll 1$. Even in weak coupling perturbation theory,
quantization imposes rather severe restrictions on possible string
theories.
In particular, all consistent string theories ($a$) live
in ten spacetime dimensions and ($b$) respect supersymmetry.
At the perturbative level, there are five such string theories, although
recent developments in nonperturbative string theory show that these
five theories are in fact perturbations around different vacua of a
single theory, whose
structure is only incompletely understood at this point
(for a review of string dualities, see Reference~\cite{asen}).

Remarkably, in all these theories there is a set of exactly massless
modes that describe the very-low-energy behavior of the
theory.  String theories have the potential to provide a unified
theory of all interactions and matter. The most common scenario for
this is to choose $l_s$ to be of the order of the Planck scale, although
there have been recent suggestions that this length  scale can be considerably
longer without contradicting known experimental facts \cite{large}.
The massless modes then describe the observed low-energy world. Of
course, to describe the real world, most of
these modes must acquire a mass, typically much smaller  than $1/l_s$.

It turns out that the massless modes of open strings are gauge
fields. The lowest state of an open string carries one quantum of the
lowest vibration mode of the
string with a polarization  $i$; this gives the gauge boson $A_i$.
The  effective low-energy field theory is a
supersymmetric Yang-Mills theory.  The closed string can carry
traveling waves both clockwise and counterclockwise along its length.
In closed string theories, the state with one quantum of the lowest
harmonic in each direction is a massless spin-2 particle, which is in
fact the graviton: If the transverse directions of the vibrations
are $i$ and $j$, then we get
the graviton $h_{ij}$. The low-energy
limits of closed string theories
thus contain gravity and are
supersymmetric extensions of general relativity---supergravity.
However, unlike these local theories of gravity, which are not
renormalizable, string theory yields a {\it finite} theory of
gravity---essentially due to the extended nature of the string.

\subsection{Kaluza-Klein Mechanism}

How can such theories in ten dimensions describe our 4-dimensional
world? The point is that
all of the dimensions need not be infinitely extended---some
of them can be compact.
Consider, for example, the simplest situation, in
which the 10-dimensional spacetime is flat and the
``internal'' 6-dimensional space is a 6-torus $T^6$
with (periodic) coordinates $y^i$, which we choose to be all of the same
period: $0 < y^i < 2\pi R$. If $x^\mu$ denotes
 the coordinates
of the noncompact 4-dimensional spacetime, we can write a scalar
field $\phi(x,y)$ as
\ben
\phi(x,y) = \sum_{n_i} \phi_{n_i} (x) ~\displaystyle{e^{i{n_i y^i \over R}}},
\label{eq:sone}
\een
where $n_i$ denotes
the six components of integer-valued momenta ${\vec n}$
along the internal directions.
When, for example, $\phi (x,y)$ is a massless field satisfying the
standard Klein-Gordon equation $(\nabla_x^2 + \nabla_y^2)\phi = 0$,
it is clear from Equation~\ref{eq:sone} that the field $\phi_{n_i} (x)$
has  (in four dimensions) a mass $m_{{\vec n}}$ given by $m_{{\vec n}} = |n|/R$.
Thus, a single field in higher dimensions becomes an infinite number
of fields in the noncompact world.
For energies much lower than $1/R$, only
 the ${\vec n} = 0$ mode
can be excited. For other kinds
of internal manifolds, the essential physics is the same.
Now, however, we have more complicated wavefunctions on the internal space.
What is rather nontrivial is that when one applies the same mechanism
to the spacetime metric, the effective lower-dimensional world
contains a metric field as well as vector gauge fields and scalar
matter fields.

\subsection{11-Dimensional and 10-Dimensional Supergravities}

Before the advent of strings as a theory of quantum gravity, there was
an attempt to control loop divergences in gravity by making the
theory supersymmetric. The greater the number of supersymmetries, the
better was the control of divergences. But in four dimensions, the maximal
number of supersymmetries is eight; more supersymmetries would force
the theory to have fields of spin higher than 2 in the graviton
supermultiplet, which leads to inconsistencies at the level of
interactions. Such $D=4, N=8$ supersymmetric theories appear
complicated but can be obtained in a simple way from a $D=11, N=1$
theory or a $D=10, N=2$ theory via the process of dimensional reduction
explained above. The gravity multiplet in the higher-dimensional
theory gives gravity as well as matter fields after dimensional
reduction to lower dimensions, with specific interactions between all
the fields.

The bosonic part of 
11-dimensional
supergravity consists of the
metric $g_{MN}$ and a 3-form gauge field $A_{MNP}$ with an action
\ben
S_{11} = {1 \over (2\pi)^8 l_p^9}[\int d^{11}x \sqrt{-g}[ R - {1\over 48}
F_{MNPQ} F^{MNPQ}] +
  {1\over 6}\int d^{11}x ~A
\wedge F
\wedge F]],
\label{eq:hone}
\een
where $R$ is the Ricci scalar and $F_{MNPQ}$ is the field strength of
$A_{MNP}$. $l_p$ denotes the 11-dimensional Planck length so that the
11-dimensional Newton's constant is $G_{11} = l_p^9$.
This theory has no free dimensionless parameter. There is only one
scale, $l_p$. Now consider compactifying one of the directions, say
$x^{10}$. The line interval may be written as
\ben
ds^2 = e^{-{\phi \over 6}}g_{\mu\nu}dx^\mu dx^\nu +
e^{{4\phi \over 3}}(dx^{11} - A_\mu dx^\mu)^2.
\label{eq:htwo}
\een
In Equation \ref{eq:htwo}, the indices $\mu,\nu$ run over the values $0 \cdots 9$.
The various components of the 11-dimensional
metric have been written in terms of a
10-dimensional metric, a field $A_\mu$ and a
field $\phi$. Clearly, from the point of view of the 10-dimensional
spacetime, $A_\mu \sim G_{\mu,10}$ is a vector and $\phi = {3\over 4}
{\rm log}~G_{10,10}$ is a scalar. In a similar way, the 3-form gauge
field splits into a rank-2 gauge field and a rank-3 gauge field
in ten dimensions, $A_{\mu\nu,10} \rightarrow B_{\mu\nu}$ and
$A_{\mu\nu\lambda} \rightarrow C_{\mu\nu\lambda}$.
The field $A_\mu$ behaves as a $U(1)$ gauge field. The bosonic massless fields
are thus
\begin{enumerate}
\item The metric $g_{\mu\nu}$; \\
\item A real scalar, the dilaton $\phi$; \\
\item A vector gauge field $A_\mu$ with field strength $F_{\mu\nu}$; \\
\item A rank-2 antisymmetric tensor gauge field $B_{\mu\nu}$ with field
strength $F_{\mu\nu\lambda}$; \\
\item A rank-3 antisymmetric tensor gauge field $C_{\mu\nu\lambda}$
with field strength $F_{\mu\nu\lambda\rho}$.
\end{enumerate}

At low energies, all
the fields in the action (Equation~\ref{eq:hone}) are independent of $x^{10}$
and the 10-dimensional action is
%
%
\bea
& S = {1 \over (2\pi)^7 g^2 l_s^8}\int d^{10} x
{\sqrt{g}}(R -  & {1\over 2}(\nabla \phi)^2 - {1\over 12}e^{-\phi}
F_{\mu\nu\alpha}F^{\mu\nu\alpha}
     - {1\over 4} e^{{3\phi\over 2}}
F_{\mu\nu}F^{\mu\nu} \nn \\
& & + {1\over 48}e^{\phi/2}F_{\mu\nu\alpha\beta}
F^{\mu\nu\alpha\beta}) + \cdots
\label{eq:hsix}
\eea
where the ellipsis denotes the terms that come from the dimensional
reduction of the last term in Equation~\ref{eq:hone}.
$F$ denotes the field strength of the appropriate gauge field.
The action (Equation~\ref{eq:hsix}) is precisely the bosonic part of the action
of Type IIA supergravity in ten dimensions. The scalar field $\phi$
is called a dilaton and plays a special role in this theory. Its
expectation value is related to the string coupling
\ben
g_s = {\rm exp}~(\langle \phi \rangle).
\label{eq:hseven}
\een
The overall factor in Equation~\ref{eq:hsix} follows from the fact that
the 11-dimensional measure in Equation~\ref{eq:hone} is related to the
10-dimensional measure by a factor of the radius of $x^{10}$ (which is
$R$),
giving
\ben
{2\pi R\over (2\pi)^8 l_p^9} = {1 \over (2\pi)^7 g^2 l_s^8},
\label{eq:height}
\een
which defines the string length $l_s$.
${}$From the 11-dimensional
metric, it is clear that $R = g^{2/3} l_p$,
so that Equation~\ref{eq:height} gives $l_p = g^{1/3} l_s$.

The 10-dimensional metric $g_{\mu\nu}$ used in Equations~\ref{eq:htwo} and
\ref{eq:hsix} is called the ``Einstein frame'' metric
because the
Einstein-Hilbert term in Equation~\ref{eq:hsix} is canonical. Other
metrics used in string
theory, most notably the ``string frame'' metric, differ from this by conformal transformations. In this
article we always use the Einstein frame metric.

Although we have given the explicit formulae for
dimensional reduction of
the bosonic sector of the theory, the
fermionic sector can be treated similarly. There are two types of
gravitinos---fermionic partners of the graviton. One of them has
positive 10-dimensional chirality whereas the other has negative
chirality. The resulting theory is thus nonchiral.

There is another supergravity in ten dimensions, Type IIB
supergravity. This cannot be obtained from D=11
supergravity by
dimensional reduction. The bosonic fields of this theory are \\
\begin{enumerate}
\item The metric $g_{\mu\nu}$ \\
\item Two real scalars : the dilaton $\phi$ and the axion $\chi$ \\
\item Two sets of rank-2 antisymmetric tensor gauge fields:
$B_{\mu\nu}$ and $B'_{\mu\nu}$ with field strengths $H_{\mu\nu\lambda}$
and $H'_{\mu\nu\lambda}$ \\
\item A rank-4 gauge field $D_{\mu\nu\lambda\rho}$ with a
self-dual field strength $F_{\mu\nu\alpha\beta\delta}$.
\end{enumerate}
Both the gravitinos of this theory have the same chirality.

Because of the self-duality constraint on the 5-form
field strength, it is not possible to write down the action for
Type
IIB supergravity,
although the equations of motion make perfect sense. If, however, we put the
5-form
field strength to zero, we have a local action given by
\bea
&S = {1 \over (2\pi)^7 g^2 l_s^8}\int d^{10} x
[{\sqrt{g}}(R - & {1\over 2}(\nabla \phi)^2
-{1\over 12}e^{-\phi}
H_{\mu\nu\alpha}H^{\mu\nu\alpha} \nn \\
& &- {1\over 12}e^{\phi}
(H'_{\mu\nu\alpha}- \chi H_{\mu\nu\alpha})(H'^{\mu\nu\alpha}
- \chi H^{\mu\nu\alpha})].\nn \\
\label{eq:hsixa}
\eea
Of course, these supergravities cannot be consistently quantized, since
they are not renormalizable. However, they are the low-energy limits
of string theories, called the Type IIA and Type IIB string.

\section{BRANES IN SUPERGRAVITY AND STRING THEORY}

Although string theory removes ultraviolet divergences leading to a
finite theory of gravity, such features as the necessity of ten dimensions
and the presence of an infinite tower of modes above the massless
graviton made it unpalatable to many physicists. Furthermore, some find the
change from a pointlike particle to a string
somewhat arbitrary---if
we accept
strings, then why not extended objects of other
dimensionalities?

Over the past few years, as nonperturbative string theory has
developed, it has been realized that the features of string
theory are actually very natural and also perhaps essential to any
correct theory of quantum gravity. A crucial ingredient in this new
insight is the fact that higher-dimensional extended
objects---branes---are present in the spectrum of string theory.

\subsection{Branes in Supergravity}

A closer look at even supergravity theories leads to the
observation that the existence of extended objects is
natural within those theories (and in fact turns out to be
essential to completing them to unitary theories
at the quantum level).

Consider the case of 11-dimensional supergravity. The supercharge
$Q_\alpha$ is a spinor, with $\alpha=1\dots 32$. The anticommutator of
two supercharge components should lead to a translation, so we write
$$\{Q_\alpha, Q_\beta\}=(\Gamma^A C)_{\alpha\beta}P_A,$$ where $C$ is
the charge conjugation matrix.  Because the anticommutator is symmetric
in $\alpha, \beta$, we find that there are $(32\times 33)/2=528$
objects on the
left-hand side
of this equation, but only $11$ objects (the $P_A$)
on the
right-hand side.
If we write down all the possible terms on the right
that are allowed by Lorentz symmetry, then we find \cite{townsend}

\ben
\{Q_\alpha, Q_\beta\}=(\Gamma^A C)_{\alpha\beta}P_A+(\Gamma^A
\Gamma^B C)_{\alpha\beta}Z_{AB}+(\Gamma^A \Gamma^B\Gamma^C
\Gamma^D\Gamma^E C)_{\alpha\beta}Z_{ABCDE},
\label{eq:stwo}
\een
where the $Z$ are
totally antisymmetric.  The number of $Z_{AB}$ is ${}^{11}C_2=55$,
whereas the number of $Z_{ABCDE}$ is ${}^{11}C_5=478$, and now we have a
total of $528$ objects on the right, in agreement with the number on the
left.

Although, for example, $ P_1\ne 0$ implies that the configuration has
momentum in direction $X^1$, what is the interpretation of $ Z_{12}\ne
0$? It turns out that this can be interpreted as the presence of a
``sheetlike'' charged object stretched along the directions $X^1, X^2$.  It is
then logical to postulate that there exists in the theory a
2-dimensional fundamental object (the 2-brane). Similarly, the
charge $Z_{ABCDE}$ corresponds to a 5-brane in the theory. The 2-brane
has a $2+1=3$-dimensional world volume and couples naturally to the
3-form gauge field present in 11-dimensional supergravity, just as a
particle with 1-dimensional world line couples to a 1-form gauge field
as $\int A_\mu dx^\mu$. The 5-brane is easily seen to be the magnetic
dual to the 2-brane, and it couples to the 6-form that is dual to the
3-form gauge field in 11 dimensions.

Thus,
it is natural to include some specific extended
objects in the quantization of 11-dimensional supergravity. But how does this
relate to string theory, which lives in ten dimensions? Let us
compactify the 11-dimensional spacetime on a small circle, thus obtaining
10-dimensional
noncompact spacetime. Then, if we let the 2-brane wrap this small
circle, we get what looks like a string in
ten dimensions. This is exactly
the Type IIA string that had been quantized by the string theorists!
The size of the small compact circle turns out to be the coupling
constant of the string.

We can also choose not to wrap the 2-brane on the small circle, in
which case there should be a two-dimensional extended object in
Type IIA
string theory. Such an object is indeed present---it is one of the
D-branes shown to exist in string theory by Polchinski
\cite{polchinski}. Similarly, we may wrap the 5-brane on the small
circle, getting a 4-dimensional D-brane in string theory, or leave it
unwrapped, getting a solitonic 5-brane, which is also known to exist in
the theory.

Thus, one is forced to a unique set of extended objects in the theory,
with specified interactions between them---in fact, there is no freedom
to add or remove any object,
nor to change any couplings.

\subsection{BPS States }

A very important property of such branes is that when they are in
an unexcited state, they preserve some of the supersymmetries of the
system, and are thus Bogomolny-Prasad-Sommerfield saturated
(BPS) states. Let us see in a simple context what a BPS state is.

Consider first a theory with a single supercharge $Q = Q^\dagger$:
\begin{equation}
{\{Q,Q\}=2Q^2=2H},
\label{eq:twotwo}
\end{equation}
where $H$ is the Hamiltonian. These relations show that the
energy of any state cannot be negative. If
\begin{equation}
{H|\psi \rangle =E|\psi \rangle}
\label{eq:twothree}
\end{equation}
then
\begin{equation}
{E= \langle \psi|H|\psi \rangle= \langle\psi|Q^2|\psi \rangle= \langle Q\psi|Q\psi \rangle~~\ge 0},
\label{eq:twothreep}
\end{equation}
where the equality holds in the last step if and only if
\begin{equation}
{Q|\psi \rangle=0},
\label{eq:twofour}
\end{equation}
that is, the state is supersymmetric. Nonsupersymmetric states occur
in a ``multiplet'' containing a bosonic state $|B \rangle$ and a fermionic
state $|F \rangle$ of the same energy:
\begin{equation}
{Q|B \rangle \equiv |F \rangle, ~~~Q|F \rangle =Q^2|B>=E|B \rangle}.
\label{eq:twofive}
\end{equation}
Supersymmetric states have $E=0$ and need not be so paired.

Now suppose there are two such supersymmetries:
\begin{equation}
{Q_1^\dagger = Q_1, ~~~Q_2^\dagger = Q_2,~~~Q_1^2=H, ~~~
Q_2^2=H,~~~\{Q_1,Q_2\}=Z},
\label{eq:twosix}
\end{equation}
where $Z$ is a ``charge''; it will take a $c$ number value on the states
that we consider below. In a spirit similar to the calculations above,
we can now conclude
\begin{equation}
{0\le~~ \langle \psi|(Q_1\pm Q_2)^2|\psi \rangle =2E \pm 2Z}.
\label{eq:twoseven}
\end{equation}
This implies that
\begin{equation}
{E\ge |Z|},
\label{eq:twoeight}
\end{equation}
with equality holding if and only if
\begin{equation}
{(Q_1-Q_2)|\psi \rangle =0,~~~ or~~~(Q_1+Q_2)|\psi \rangle =0}.
\label{eq:twonine}
\end{equation}
Now we have three kinds of states:
\begin{enumerate}
\item{} States with $Q_1|\psi \rangle =Q_2|\psi \rangle =0$. These
have $E=0$ and do not fall into a multiplet. By Equation~\ref{eq:twoseven},
they also have $Z|\psi \rangle =0$, so they carry no charge.
\item{} States not in category 1, but satisfying Equation~\ref{eq:twonine}.
For concreteness, take the case $(Q_1-Q_2)|\psi \rangle =0$.
These states fall
into a ``short multiplet'' described by, say, the basis $\{|\psi \rangle ,~
Q_1|\psi \rangle \}$. Note that
\begin{equation}
{Q_2|\psi \rangle =Q_1|\psi \rangle , ~~~Q_2Q_1|\psi \rangle =Q_1^2|\psi \rangle =E|\psi \rangle },
\label{eq:twoninea}
\end{equation}
so that we have no more linearly independent states in the
multiplet. Such states satisfy
\begin{equation}
{E=|Z| > 0}
\label{twoten}
\end{equation}
and are called BPS states. Note that by Equation~\ref{eq:twoseven}, the state with
$Z \rangle 0$ satisfies $(Q_1-Q_2)|\psi \rangle =0$ while the state with $Z<0$
satisfies $(Q_1+Q_2)|\psi \rangle =0$.
\item{} States that are not annihilated by any linear combination of
$Q_1, Q_2$. These form a ``long multiplet'' $\{|\psi \rangle , ~Q_1|\psi \rangle ,~
Q_2|\psi \rangle,~ Q_2Q_1|\psi \rangle \}$. They must have $E>|Z|>0$.
\end{enumerate}

In the above discussion, we have regarded the BPS states as states of a
quantum system, but a similar analysis
applies to classical solutions. In 10-dimensional supergravity, the branes mentioned above  appear as
classical solutions of the equations of motion, typically with sources.
They are massive solitonlike objects and therefore produce
gravitational fields. Apart from that, they produce the p-form gauge
fields to which they couple and, in general, a nontrivial dilaton.
A brane in a general configuration would break all the
supersymmetries of the theory. However, for special configurations---corresponding to ``unexcited branes''---there are solutions which
retain some of the supersymmetries. These are  BPS saturated
solutions, and since they have the maximal charge for a given mass,
they are stable objects. We use such branes below in constructing
black holes.

\subsection{The Type IIB Theory}

Instead of using the 11-dimensional algebra, we could
have used the 10-dimensional algebra and arrived at the
same conclusions. In a similar fashion, the existence of BPS branes
in Type
IIB supergravity follows from the corresponding algebra. For
each antisymmetric tensor field present in the spectrum,
there is a corresponding BPS brane. Thus we have

\begin{enumerate}
\item{} D($-1$)-brane, or D-instantons,
carrying charge under the axion
field $\chi$;
\item{} NS1-brane, or elementary string,
carrying electric charge under $B_{\mu\nu}$;
\item{} D1-brane, carrying electric charge under $B'_{\mu\nu}$;
\item{} 3-brane, carrying charge under $D_{\mu\nu\rho\lambda}$;
\item{} NS5-brane, carrying magnetic charge under $B_{\mu\nu}$;
\item{} D5-brane, carrying magnetic charge under $B'_{\mu\nu}$;
\item{} D7-brane, the dual of the D($-1$) brane.
\end{enumerate}
We have denoted some of the branes as D-branes. These play a special
role in string theory, as explained below.

\subsection{D-Branes in String Theory}

Consider the low-energy action of the supergravity theories. We have
remarked above that the equations of motion for the massless fields
admit solitonic solutions, where the solitons are not in general
pointlike but could be $p$-dimensional sheets in space [thus having
$(p+1)$-dimensional world volumes in spacetime]. In each of these cases,
the soliton involves the gravitational field and some other
$(p+1)$-form field in
the supergravity multiplet so that the final solution carries a charge
under this $(p+1)$-form
gauge field. In fact, in appropriate units, this soliton is seen to
have a mass equal to its
charge and is thus an object satisfying the BPS bound of
supersymmetric theories.
This fact implies that the soliton is a stable construct in the
theory.
The possible values of $p$ are determined entirely by the properties
of fermions in the Type II theory. It turns out that for
Type IIA, $p$ must
be even ($p = 0,2,4,6$), whereas for Type
IIB $p$ must be odd ($p = -1,1,3,5,7$).
Recalling the massless spectrum of these theories, we find that for
each antisymmetric tensor field there is a brane that couples to it.
Because the brane is not pointlike but is an extended object, we can
easily see that there will be low-energy excitations of this soliton
in which its world sheet suffers small transverse displacements that vary
along the brane (in other words, the brane carries waves corresponding
to transverse vibrations).  The low-energy action is thus expected to
be the tension of the brane times its area.  For a single 1-brane, for
instance, the action for long-wavelength deformations is
\ben
S = {T_1 \over 2}\int d^2\xi^\alpha {\sqrt{{\rm{det}}(g_{\alpha\beta})}},
\label{eq:sseven}
\een
where
$\xi^\alpha$, with $\alpha=1,2$, denotes
an arbitrary coordinate sytem
on the D1-brane world sheet and $g_{\alpha\beta}$ denotes the
induced metric on the brane,
\ben
g_{\alpha\beta} = \partial_\alpha X^\mu \partial_\beta X^\nu
\eta_{\mu\nu}.
\label{eq:seight}
\een
The $X^\mu,$
with $\mu = 0, \cdots, 9$,
denotes the coordinates of a point on the brane. This action
is invariant under arbitrary transformations of the coordinates $\xi^\alpha$
on the brane. To make contact with
to the picture discussed above, it is best to
work in a static gauge by choosing $\xi^1 = X^0$ and $\xi^2 = X^9$. The
induced metric (Equation~\ref{eq:seight}) then becomes
\ben
g_{\alpha\beta} = \eta_{\alpha\beta}+ \partial_\alpha \phi^i \partial_\beta
\phi^j \delta_{ij},
\een
where $\phi^i=X^i$ with $i = 1,\cdots 8$, are the remaining fields.
The determinant in Equation~\ref{eq:sseven} may be then expanded in powers of
$\partial \phi^i$. The lowest-order term is just the free kinetic energy
term for the eight scalar fields $\phi^i$.

It is now straightforward to extend the above discussion for
higher-dimensional branes.
Now we can have both transverse and longitudinal oscillations of the
brane. For a $p$-brane we have $(9-p)$ transverse coordinates,
labeled by $I = 1, \cdots (9-p)$,
and hence as many scalar fields on the $(p+1)$-dimensional brane
world volume, $\phi^i$. It turns out that the longitudinal waves are
carried by a $U(1)$ {\it gauge} field $A_\alpha$ with the index
$\alpha = 0,1,\cdots p$ ranging over the world volume directions. The
generalization of  Equation~\ref{eq:sseven} [called the
Dirac-Born-Infeld
(DBI) action] is
\ben
S = {T_p \over 2}\int d^{p+1}\xi {\sqrt{{\rm{det}}(g_{\alpha\beta}
+ F_{\alpha\beta})}},
\een
where $F_{\alpha\beta}$ is the gauge field strength. Once again, one can
choose a static gauge and relate the fields directly to a string
theory description. The low-energy expansion of the action then leads
to electrodynamics in $(p+1)$ dimensions with some neutral scalars
and fermions.

In the above description, we had obtained the branes as classical
solutions of the supergravity fields; this is the analog of
describing a point charge by its
classical electromagnetic potential. Such a description should apply
to a  collection of a large number of fundamental branes all placed
at the same location. But
we would like to obtain also the microscopic quantum physics of a
single brane. How do we see such an object in string theory?

The mass per unit volume of a D-brane in string units is $1/g$, where
$g$ is the string coupling. So the brane would not be seen as a
perturbative object at weak coupling. But the excitations of the
brane will still be low-energy modes, and these should be seen in weakly
coupled string theory.  In fact, when we quantize a free string we have
two choices: to consider open strings or closed strings. If we have an
open string then we need boundary conditions at the ends of
the string that do not allow the energy of vibration of the string to
flow off the
end. There are two possibilities:
to let the ends move at the speed of light, which corresponds to Neumann
(N) boundary conditions, or to fix the end, which corresponds
to Dirichlet
(D) boundary conditions. Of course we can choose different
types of conditions for different directions of motion in
spacetime. If the ends of the open strings are free to move along the
directions $\xi^\alpha$ but are fixed in the other directions $X^i$,
then the ends are constrained
to lie along a $p$-dimensional
surface that we may identify with a $p$-brane, and such open strings
describe the excitations of the $p$-brane
\cite{polchinski}.  Because these branes were
discovered through their excitations, which were in turn a consequence
of D-type boundary conditions on the open strings, the branes are
themselves called D-branes. (For a review of properties of D-branes,
see Reference~\cite{tasilectures}.)

An interesting effect occurs when two  such branes are brought close
to each other.
The open strings that begin and end
on the first brane will describe excitations of the first brane,
and those that begin and end on the second brane will
describe excitations of the second brane. But, as shown in Figure~2, an open string can also
begin on the first brane and end on the second,
or begin on the second and end on the first, and this gives two
additional possibilities for the excitation of the
system. These four possibilities can in fact be encoded into a
$2\times 2$ matrix, with the $(ij)$ element of the
matrix given by open strings that begin on the $i$th brane and end on
the $j$th brane. This structure immediately
extends to the case where $N$ branes approach each other.

\begin{figure}
\epsfysize=6cm \epsfbox{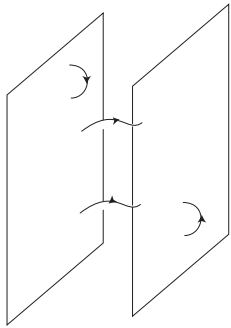}
\caption{Open strings ending on D-branes.}
\end{figure}

The low-energy limits
 of open string theories are generally gauge
theories.  Indeed the low-energy worldbrane theory of a collection of
$N$ parallel D$p$-branes
turns out to be described by a {\it
non-Abelian}
$U(N)$ gauge theory in $(p+1)$ dimensions
\cite{wittendbranes}. The $N^2$
gauge fields are best written as
an $N \times N$ matrix
$A^\alpha_{ab},$ where $a,b = 1\cdots N$. Similarly, there are
$N^2$ scalar fields
$\phi^i_{ab}$, which transform according to the adjoint representation
of $U(N)$. The coupling constant of the theory is
\ben
g_{{\mathrm YM}}^2 \sim g_s (l_s)^{p-3}.
\een
Each of these fields has its
corresponding fermionic partner.
Since BPS states break half the supersymmetries of the original
theory, we have a supersymmetric Yang-Mills theory with 16
supercharges.

The potential for such a theory turns out to be ${\rm
Tr}([\phi^i,\phi^j])^2$, so that in the ground state one can always
choose a gauge so that all the $\phi^i$s are diagonal.
In fact, the
diagonal entries in the matrix $\phi^i_{ab}$ denote the locations of the
branes. Thus ${1\over 2}(\phi^i_{aa} + \phi^i_{bb})$ denotes
 the center-of-mass transverse coordinate
of the pair of branes labeled by $a$
and $b$, while $d^i_{ab} = (\phi^i_{aa} - \phi^i_{bb})$ denotes
 the
separation along direction $i$. A nonzero expectation value for
$d^i_{ab}$ means that the gauge symmetry is spontaneously broken.

Generically, the gauge group is broken to $[U(1)]^N$---this is the
situation when all the branes are separated from each other, i.e.\  all
the $d^i_{ab}$ are nonzero. However, when some number, say $M$, of the
branes are coincident, the corresponding $d^i_{ab}$ are zero for $a,b
= 1 \cdots M$,
resulting in
an enhanced unbroken symmetry, $U(M) \times
[U(1)]^{N-M}$.

The analog of the DBI action for a collection of $N$ branes is not
completely known at
present, although there are some proposals (for
discussion see Reference~\cite{tseytlin} and references therein).  However,
the low-energy action is that of a standard supersymmetric Yang-Mills
theory in $p+1$ dimensions.  In fact, this is the dimensional reduction
of the 10-dimensional supersymmetric Yang-Mills theory to $p+1$
dimensions. The latter has no scalar fields but has $10$ components
of the gauge field, $A_\mu,$ where $\mu = 0,\cdots, 9$, each of which is a $N
\times N$ matrix. Under dimensional reduction, the components parallel
to the brane, i.e.\ $\mu = 0,(10-p),\cdots, 9$, remain gauge fields,
whereas the components transverse to the brane with $\mu = 1,\cdots
(9-p)$ are scalars on the world volume and are renamed as $\phi^i$.

The ground states of
the above D-branes are BPS states
states, which means that they are stable.
Recently,
other kinds of non-BPS and unstable D-branes have been constructed in
string theory \cite{sennonbps}. In this article, however,
we restrict ourselves to BPS branes and their excitations.

Not all the branes that were listed for Type IIA and Type IIB theory
are D-branes. Consider the Type IIA theory.  It arises from a
dimensional reduction of 11-dimensional supergravity on a circle. The
11-dimensional
theory has 5-branes and 2-branes, and the 2-brane can end on the
5-brane. If we wrap on the compact circle both the 5-brane  and the
2-brane
that ends
on it, then in the Type IIA theory we get a D4-brane and
an open string ending on the D4-brane. But if we do not wrap the
5-brane on the circle, and thus get a 5-brane in the Type IIA theory,
then the open string cannot end on this brane, since there is no
corresponding picture of such an endpoint in
11 dimensions. The physics of the
5-brane in Type IIA theory is an interesting one, but we will not
discuss it further. Much more is known about the physics of D-branes
in Type
IIA and IIB theories, since they can be studied through
perturbative open string theory.

Branes of different kinds can
also form bound states, and for specific instances these can be
threshold bound states.
A useful example is a bound state of $Q_1$
D1-branes and
$Q_5$ D5-branes.
When these branes are not excited, they are in a threshold bound state.
The open strings that describe this system
are ($a$) $(1,1)$ strings with both endpoints on any of the D1-branes;
($b$) $(5,5)$ strings with both endpoints on D5-branes; and ($c$)
$(1,5)$ and
$(5,1)$ strings with one endpoint on any of the D1-branes and the
other endpoint on one of the D5-branes.

\subsection{Duality}

A remarkable feature of string theory is the large group of
symmetries called dualities
(see Reference~\cite{asen} for review). Consider
Type IIA
theory compactified on a circle of radius $R$. We can take
a graviton
propagating along this compact direction; its energy spectrum would
have the form $|n_p|/R$. But we can also wind an elementary string on
this circle; the spectrum here would be $2\pi T_s|n_w| R$, where $n_w$
is the winding number of the string and $T_s$ its tension. We note
that if we replace $R\rightarrow {1/2\pi T_s R}$, then
the energies of the above two sets of states
are simply
interchanged.  In
fact this map, called T-duality,  is an exact symmetry of string
theory; it interchanges winding and momentum modes, and  because
of an effect on 
fermions, it also turns a Type IIA theory into Type IIB and vice versa.

The Type IIB theory also has another symmetry called S-duality,
there the string coupling $g_s$ goes to ${1/g_s}$. At the same
time,
the role of the elementary string is interchanged with the
D1-brane. Such a duality, which relates weak coupling to strong
coupling while interchanging fundamental quanta with solitonic objects,
is a realization of the duality suggested for field theory by Montonen
\& Olive \cite{olive}. The combination of S- and T-dualities generates a
larger group, the U-duality group.

There are other dualities, such as those that relate
Type IIA theory to heterotic string theory, and those that relate these theories to the
theory of unoriented strings.
In this article, we do not use the idea of dualities
directly, but we note that any black hole that we construct by
using branes is related by duality maps to a large class of similar holes that have the same
physics, so the physics
obtained is much more universal than it may at first appear.

\section{BLACK HOLE ENTROPY IN STRING THEORY: THE FUNDAMENTAL
STRING}

String theory is a quantum theory of gravity. Thus, black holes
should appear in this theory as excited quantum states.
An idea of Susskind \cite{susskind} has proved very useful in the
study of black holes. Because the coupling in the theory is not a
constant but a variable field,
we can study a state of the theory
at weak coupling, where we can use our knowledge of string
theory. Thus we may compute the ``entropy'' of the state, which would be
the logarithm of the number of states with the same mass and
charges. Now imagine the coupling to be tuned to strong values. Then
the gravitational coupling also increases, and the object must become
a black hole with a large radius. For this black hole we can compute the
Bekenstein entropy from (Equation~\ref{eq:two}), and ask if the microscopic
computation agrees with the Bekenstein entropy.

For such a calculation to make sense, we must have some
assurance that the density of states would not shift when we change
the coupling. This is where BPS states come in. We have
shown above that the masses of BPS saturated states are indeed
determined once we know their charges, which are simply their winding
numbers on cycles of the compact space.
Thus, for such states, we
may calculate the degeneracy of states at weak coupling and, since the
degeneracy can be predicted also at strong coupling,  compare the
result with the Bekenstein-Hawking entropy of the corresponding
black hole.  Such states give, at strong coupling, black holes that are
``extremal''---they have the minimal mass for their charge if we
require that the metric does not have a naked singularity.

The extended objects discussed in the previous section have played
an important role in understanding black holes in string theory.

An example of such an object is a fundamental string in Type IIA or
IIB string theory.  Let some of the directions of the 10-dimensional
spacetime
be compactified on small circles. Take a fundamental string and wrap
it $n_1$ times around one of these circles, say along the $9$
direction, which has a radius $R$. This will produce a rank-2 NS
field $B_{09}$ with a charge $n_1$, which is a ``winding charge.''
The energy of the state is $E =
2\pi n_1 T_s R$, which saturates the BPS bound.  From the point of view
of the noncompact directions, this looks like a massive point object
carrying electric charge under the gauge field that results
from the
dimensional reduction of the rank-2 field.
From the microscopic viewpoint, the state
of such a string is unique (it does have a 256-fold degeneracy due to
supersymmetry, but we can ignore this---it is not a number that grows
with $n_1$). Thus, the microscopic entropy is zero.

If we increase the
coupling, we expect a charged black hole.
Furthermore, this is an extremal black hole,
since for a given charge $n_1$ this has the lowest allowed mass given
by the energy given above.  However, this black hole turns out to have
a vanishing horizon area.
One way to understand this is to note that the
tension of the string ``pinches'' the circle where the string was
wrapped.
Thus, entropy is zero from both the microscopic and the black hole viewpoints,
which is consistent but not really
interesting.

To prevent this pinching, we can put some momentum along the string,
which amounts to having traveling waves move along the string. The
momentum modes have an energy that goes as
$1/R$, so now this circle
attains a finite size. If we put waves that are moving only in one of
the directions,
we still have a BPS saturated state, but with a further
half of the supersymmetries broken. The total energy of the state is
now given by $E = 2\pi n_1 T_s + n_2/R$, where the second term is now
the contribution from the momentum waves. Because of the winding, the
effective length of the string is $L_{{\mathrm eff}} = 2\pi n_1 R$, so that the
momentum may be written as $P = (2\pi n_1 n_2)/L_{{\mathrm eff}}$. Thus, for
given values of $n_1$ and $n_2$, a large number of states
have the same energy. These correspond
to the various ways one can
get the oscillator level $n_0= n_1n_2$.
In addition,
it is necessary to consider the fact that there are eight
possible polarizations, and
there are fermionic oscillators as well. The resulting degeneracy of
states has been known since the early days of string theory.  For
large $n_1$ and $n_2$, the number of states asymptotes
to
\ben
n(n_1,n_2) \sim {\rm{exp}}[2\sqrt{2}\sqrt{n_1n_2}].
\label{eq:bhone}
\een

From the viewpoint of the noncompact directions, we have an
object with two quantized charges $n_1$ and $n_2$.  The charge $n_1$
is due to the winding of the string, as mentioned above. The second
charge is due to the presence of momentum, which results in a term
in the 10-dimensional metric proportional to $dx^9 dx^0$. Then, by
the Kaluza-Klein mechanism explained in the previous section, this is
equivalent to a gauge field $A'_0$ from the viewpoint of the
noncompact world.  The corresponding charge is $n_2$ and is an
integer since the momentum in the compact direction is
quantized.  At strong coupling, this object is described by an extremal
black hole solution with two charges. The identification of such
fundamental string states with the corresponding classical solution
was proposed by Dabholkar \& Harvey
\cite{dharvey}. The
horizon area is still zero and the curvatures are large near the
horizon.

In an important paper, Sen \cite{sen} argued that the semiclassical
entropy (for similar black holes in heterotic string theory)
is given not by the area of the event horizon but by the area
of a ``stretched horizon.'' This is defined as the location
where the curvature and local temperature reach the string scale. It
is indeed inconsistent to trust a classical gravity solution beyond
this surface, since the curvatures are much larger than the string scale
and stringy corrections to supergravity become relevant. There is
a great deal of ambiguity in defining the stretched horizon precisely.
However, Sen found that the area of the stretched horizon in
units of the gravitational constant is proportional to
$\sqrt{n_1n_2}$, which is precisely the logarithm of the degeneracy
of states given by Equation~\ref{eq:bhone}.
This was the first indication that
degeneracy of states in string theory may account for Bekenstein-Hawking
entropy. However, in this example, the precise coefficient cannot be
determined, since the definition of the stretched horizon is itself
ambiguous.

It is clear that what we need is a black hole solution that
($a$) is BPS and ($b$) has a nonzero large horizon area.
The BPS nature of
the solution would ensure that degeneracies of the corresponding string
states are the same at strong and weak couplings, thus allowing an
accurate computation of the density of states. A large
horizon would ensure that the curvatures are weak and we can therefore
trust semiclassical answers.
In that situation, a microscopic count of the states
could be compared with the Bekenstein-Hawking entropy in a regime where both calculations
are trustworthy.

It turns out that
this requires three kinds of
charges for a 5-dimensional black hole and four kinds of charges
for a 4-dimensional black hole.

\section{THE FIVE-DIMENSIONAL BLACK HOLE IN TYPE
IIB THEORY}

The simplest black holes of this type are in fact 5-dimensional
charged black holes. Extremal limits of such black holes
provided the first example
where the degeneracy of the corresponding BPS states exactly
accounted for the Bekenstein-Hawking
entropy \cite{stromingervafa}.
There are several such black holes in Type IIA and IIB theory, all related to each other by string dualities. We
describe below one such solution in Type IIB supergravity.

\subsection{The Classical Solution}

We start with 10-dimensional spacetime and compactify on a $T^5$
along $(x^5,x^6,$
$  \cdots x^9)$. The noncompact directions are then
$(x^0.\cdots x^4)$. There is a solution of Type
IIB supergravity that
represents ($a$) D5-branes wrapped around the $T^5$,
($b$) D1-branes wrapped around $x^5$, and
($c$) some momentum along $x^5$.
Finally, we perform a Kaluza-Klein reduction to the five noncompact
dimensions. The resulting metric is\footnote{We use the notation
of Horowitz et al \cite{horomalstrom}.}
\ben
ds^2 =
- [f(r)]^{-2/3} \left(1-{r_0^2 \over r^2} \right)dt^2
+ [f(r)]^{1/3} \left[{dr^2 \over 1- {r_0^2 \over r^2}}+r^2 d\Omega_3^2 \right],
\label{eq:fdone}
\een
where
\ben
f(r) \equiv \left(1 + {r_0^2\sin^2\alpha_1 \over r^2} \right)
\left(1 + {r_0^2 \sin^2 \alpha_5\over r^2} \right)
\left(1 + {r_0^2 \sinh^2 \sigma \over r^2} \right).
\label{eq:fdtwo}
\een
Here $r$ is the radial coordinate in the transverse space,
$r^2 = \sum_{i=1}^4 (x^i)^2$,
and $d\Omega_3^2$
is the line element on a unit 3-sphere $S^3$.
The solution represents a black hole with an outer horizon at $r = r_0$
and an inner horizon at $r = 0$.
The background also has nontrivial values of the dilaton $\phi$, and
there are three kinds of gauge fields:

\begin{enumerate}
\item{} A gauge field $A^{(1)}_0$, which comes from the dimensional
reduction of the rank-2 antisymmetric tensor gauge field $B'_{05}$ in 10
dimensions. This is nonzero, since we have D1-branes
along $x^5$.
\item{} A gauge field $A^{(2)}_0$, which comes from the dimensional
reduction of an off-diagonal component of the 10-dimensional
metric $g_{05}$. This is nonzero,
since there is momentum
along $x^5$.
\item{} A rank-2 gauge field $B'_{ij}$ with $i$ and $j$ lying along
the 3-sphere $S^3$. This is the dimensional reduction of a
corresponding $B'_{ij}$ in ten dimensions, since there are 5-branes
along $(x^5 \cdots x^9)$.
\end{enumerate}
The presence of these gauge fields follows in a way exactly
analogous to the dimensional reduction
from 11-dimensional supergravity discussed in Equations~\ref{eq:hone}
and \ref{eq:htwo}. The corresponding
charges $Q_1,Q_5,N$ are given by
\ben
Q_1  =  {V r_0^2 \sinh 2\alpha_1 \over 32
\pi^4 g l_s^6}~~~~
     Q_5  =  {r_0^2 \sinh 2\alpha_5 \over 2 g l_s^2}~~~~
N  =  {V R^2 \over 32 \pi^4 l_s^8 g^2}
r_0^2 \sinh 2\sigma,
\label{eq:fdfive}
\een
where $V$ is the volume of the $T^4$ in the $x^6 \cdots x^9$
directions, $R$ is the radius of the
$x^5$ circle, and $g$ is the string
coupling.

The charge $N$ comes from momentum in the $x^5$ direction. If we look
at the higher-dimensional metric before dimensional reduction, it is
straightforward to identify the Arnowitt-Deser-Misner (ADM) momentum as
\ben
P_{{\mathrm ADM}} = {N \over R}
\label{eq:fdfivea}
\een
The ADM mass of the black hole is given by
\ben
M = {RV r_0^2 \over 32\pi^4 g^2 l_s^8}[\cosh 2\alpha_1
+ \cosh 2\alpha_5 + \cosh 2\sigma].
\label{eq:fdsix}
\een

\subsection{Semiclassical Thermodynamics}

The semiclassical thermodynamic properties of the black hole
may be easily obtained from the classical solution using Equations~\ref{eq:two} and \ref{eq:nthree} and the relationship
\ben
G_5 = {4\pi^5 g^2 l_s^8 \over RV}
\label{eq:fdseven}
\een
between the 5-dimensional Newton's constant $G_5$ and
$V,R,g,l_s$.

The expressions are
\bea
S_{{\mathrm BH}} & = & {A_H \over G_5} =
{RV r_0^3 \over 8 \pi^3 l_s^8 g^2}
\cosh \alpha_1 \cosh \alpha_5 \cosh \sigma \nn \\
{\rm and}\\
T_{\mathrm H} & = &{1 \over 2\pi r_0 \cosh \alpha_1 \cosh \alpha_5
\cosh \sigma}.
\label{eq:fdnine}
\eea

\subsection{Extremal and Near-Extremal Limits}

The solution given above represents a general 5-dimensional
black hole. The extremal limit is defined by
\bea
&r_0^2 \rightarrow 0&\alpha_1,\alpha_5,\sigma \rightarrow \infty \nn \\
&Q_1,Q_5,N = {\rm fixed}.
\label{eq:fdten}
\eea
In this limit, the inner and outer horizons coincide. However, it is
clear from the above expressions that ADM mass and entropy remain
finite while the Hawking temperature vanishes.  Of particular interest
is the extremal limit of the entropy,
\ben
S_{{\mathrm extremal}} = 2\pi {\sqrt{Q_1Q_5N}}.
\label{eq:fdeleven}
\een
This is a function of the charges alone and is independent of other
parameters, such as the string coupling, volume of the compact directions,
and string length.

In the following, we are interested in a special kind of departure
from extremality. This is the regime in which $\alpha_1, \alpha_5 \gg 1$,
but $r_0$ and $\sigma$
are finite. In this case, the total ADM mass may be
written in the suggestive form
\ben
E = {R Q_1 \over g l_s^2} +
{16 \pi^4 RV Q_5 \over l_s^6} + E_L + E_R.
\label{eq:fdtwelve}
\een
Here we have defined
\ben
E_L = {N \over R} +
{VR r_0^2 e^{-2\sigma} \over 64 \pi^4 g^2 l_s^8}
~~~~{\rm and}~~~~~
E_R =   {VR r_0^2 e^{-2\sigma} \over 64 \pi^4 g^2 l_s^8},
\label{eq:fdthirteen}
\een
and used the approximations
\bea
r_1^2 & \equiv & r_0^2 \sinh^2 \alpha_1
\sim {r_0^2 \over 2} \sinh 2\alpha_1 =
{16 \pi^4 g l_s^6 Q_1 \over V}, \nn \\
r_5^2 & \equiv & r_0^2 \sinh^2 \alpha_5
\sim {r_0^2 \over 2} \sinh 2\alpha_5 =
g l_s^2 Q_5,
\label{eq:fdfourteen}
\eea
and the expressions for the charges (Equation~\ref{eq:fdfive}).
The meaning of the subscripts $L$ and $R$
will be clear soon.  In a similar
fashion, the thermodynamic quantities may be written as
\ben
     S = S_L + S_R ~~~~~~~~~~~~~{1 \over T_H} =
{1\over 2}\left({1\over T_L} + {1\over T_R}\right),
\label{eq:fdfifteen}
\een
where
\ben
S_{L,R} = { R V r_1 r_5 r_0 e^{\pm \sigma}
\over 16 \pi^3 g^2 l_s^8}~~~~~~
T_{L,R} = {r_0 e^{\pm \sigma} \over 2\pi r_1 r_5}.
\label{eq:fdsixteen}
\een
The above relations are highly suggestive.  The contribution to the
ADM mass from momentum along $x^5$ is $E_m = E_L + E_R$. In the
extremal limit $\sigma \rightarrow
\infty$, it then follows from Equations~\ref{eq:fdthirteen} and
\ref{eq:fdfivea} that $E_m \rightarrow P_{{\mathrm ADM}}$, which implies
that the waves are moving purely in one direction. This is the
origin of the subscripts $L$ and $R$---we have denoted the direction
of momentum in the extremal limit as left, $L$. For finite $\sigma$
we have both right- and left-moving waves, but
the total momentum is still $N/R$. The splitting of various
quantities into left- and right-moving parts is typical of waves
moving at the speed of light in one space dimension. In fact, using
Equations~\ref{eq:fdfifteen}, \ref{eq:fdsixteen}, and
\ref{eq:fdfive}, we can easily see that the following relations
hold:
\ben
T_i = {1\over \pi}{\sqrt{{E_i \over RQ_1Q_5}}}
= {S_i \over 2\pi^2 RQ_1Q_5},~~~~i = L,R.
\label{eq:fdseventeen}
\een
Finally, we write the expressions that relate the temperature $T$
and the entropy $S$ to the excitation energy over extremality $\Delta E
= E_m - N/R$:
\bea
\Delta E & = & {\pi^2 \over 2}(RQ_1Q_5)~T^2 \nn \\
{\rm and}~~~S & = & S_{{\mathrm extremal}}[ 1 + ({R \Delta E \over 2N})^{{1\over 2}}].
\label{eq:fdeighteen}
\eea

\subsection{Microscopic Model for the Five-Dimensional Black Hole}

This solution and the semiclassical properties described above are
valid in regions where the string frame curvatures are small compared
with
$l_s^{-2}$. If we require
this to be true for the entire region
outside the horizon, we can study the regime of validity of the
classical solution in terms of the various parameters. A short
calculation using Equations~\ref{eq:fdone}, \ref{eq:fdtwo}, and
\ref{eq:fdfive} shows that this is given by
\ben
(gQ_1), (gQ_5), (g^2 N) \gg 1.
\een
In this regime, the classical solution is a good description of an
intersecting set of D-branes in string theory.

However, in string theory, the way to obtain a state with large
D-brane charge $Q$ is to have a collection of $Q$ individual
D-branes. Thus, the charges $Q_1$ and $Q_5$ are integers, and we
have a system of $Q_1$ D1-branes and $Q_5$ D5-branes.
The third
charge is from a momentum in the $x^5$ direction equal to $N/R$.
Thus, $N$ is quantized in the microscopic picture to ensure a
single valued wave function.

In the extremal limit (Equation~\ref{eq:fdten}), these branes are in a
threshold bound state, i.e.\ a bound state with zero binding
energy. This is
readily apparent from the classical solution. The
total energy (Equation~\ref{eq:fdtwelve}) is then
seen to be the sum of
the masses of $Q_1$ D1-branes, $Q_5$ D5-branes and the total momentum
equal to $N/R$.

The low-energy theory of a collection of $Q$ D$p$-branes
is a ($p+1$)-dimensional supersymmetric gauge theory with gauge group
$U(Q)$. However, now we have a rather complicated bound state of
intersecting branes. The low-energy theory is still a gauge theory
but has additional matter fields (called hypermultiplets).  Instead
of starting from the gauge theory itself, we first present a
physical picture of the low-energy excitations.

\subsection{The Long String and Near-Extremal Entropy}

The low-energy excitations of the system become particularly
transparent in the regime where the radius of the $x^5$ circle, $R$,
is much larger than the size of the other four compact directions
$V^{1/4}$. Then the effective theory is a $(1+1)$-dimensional theory
living on $x^5$.
The modes are essentially those of the oscillations
of the D1-branes. These, however, have four rather than eight
polarizations. This is because it costs energy to pull a D1-brane away
from the D5-brane, whereas the motions of the D1-branes along the
four directions parallel to the D5-branes,
but transverse to the D1-branes
themselves (i.e.\ along $x^6 \cdots x^8$), are gapless excitations.

Even though
a system of static D1- and D5-branes is
marginally bound,
there is a nonzero binding energy whenever the D1-branes try to move
transverse to the D5-branes.  If we had a single D1-brane and a single
D5-brane, the quantized waves would be
massless particles with four
flavors. Since we have a supersymmetric theory, we have four bosons and
four fermions. When we have many D1-branes and D5-branes, it would
appear at first sight that there should be $4Q_1Q_5$ such flavors:
Each of the D1-branes can oscillate along each of the D5-branes, and
there are four polarizations for each such oscillation. This is indeed
the case if the D1-branes are all separate.

However, there are other possible configurations. These correspond to
several of the D1-branes joining up to form a longer 1-brane, which is
now multiply wound along the compact circle.
In fact, if we only have some number $n_w$ of D1-branes without anything
else wrapping a compact circle, they would also prefer
to join into
a long string, which is of length $2\pi n_w R$. This was discovered
in an analysis of the nonextremal excitations of such D1-branes
\cite{dasmathur1}.
S-duality relates D1-branes to fundamental wrapped
strings.
If one requires that the nonextremal excitations of the D1
brane are in one-to-one correspondence with the known near-extremal
excitations of the fundamental string and therefore yield the same
degeneracy of states, one concludes that the energies
of the individual quanta of oscillations of the D1 string must have
fractional momenta $p_i = n_i/(n_w R)$, which simply means that
the effective length of the string is $2\pi n_w R$.

For the situation we are discussing, 
i.e.\
D1-branes bound to D5-branes,
the entropically favorable configuration turns out to be that of a
single long string formed by the D1-branes joining up to
wind around the $x^5$ direction $Q_1Q_5$, i.e.\ an effective length of
$L_{{\mathrm eff}} = 2\pi R Q_1Q_5$
\cite{maldasusskind}. We thus arrive at a gas of four flavors
of bosons and four flavors of fermions living
on a circle of size
$L_{{\mathrm eff}}$. Since the string is relativistic, these particles are
massless. The problem is to count the number of states with a given
energy $E$ and given total momentum $P = N/R$ and see if the results
agree with semiclassical thermodynamics of black holes. This approach
to the derivation of black hole thermodynamics
comes from Callan \& Maldacena \cite{callanmalda}
and Horowitz \& Strominger \cite{horostrom},
with the important modification
of the multiply wound D-string
proposed by Maldacena \& Susskind \cite{maldasusskind}. Our treatment below follows Reference~\cite{dasmathur}.

\subsubsection{Statistical Mechanics in $1+1$ Dimensions}

At weak coupling these bosons and fermions form an ideal gas, which we
assume thermalizes. Consider the general case of such an ideal gas with
$f$ flavors of massless bosons and fermions (each polarization counted
once) living
on a circle of size $L$. For large $L$, we should be able
to treat the system in a canonical ensemble characterized by an inverse
temperature $\beta$ conjugate to the energy and a chemical potential
$\alpha$ conjugate to the momentum. If $n_r$ denotes the number of
particles with energy $e_r$ and momentum $p_r$, the partition function
${\cal Z}$ is
\ben {\cal Z} = e^h = \sum_{{\mathrm states}} {\rm exp}~\left[
-\beta\sum_r n_r e_r - \alpha\sum_r n_r p_r \right].
\label{eq:therone}
\een
Then $\alpha$ and $\beta$ are determined by requiring
\ben
E = -{\partial h \over \partial \beta}~~~{\rm and}~~~~
P = -{\partial h \over \partial \alpha}.
\label{eq:thertwo}
\een
The average number of particles $n_r$ in state $(e_r, p_r)$ is
then given by
\ben
\rho (e_r, p_r) = {1 \over e^{\beta e_r + \alpha p_r}
\pm 1},
\label{eq:therthree}
\een
where as usual the plus sign is for fermions and the minus sign is
for bosons. Finally, the entropy $S$ is given by the standard thermodynamic
relation
\ben
S = h + \alpha P + \beta E.
\label{eq:therfour}
\een
The above quantities may be easily evaluated:

\bea
P & = & {fL\pi\over 8} \left[{1\over (\beta + \alpha)^2}
-{1 \over (\beta - \alpha)^2}\right], \nn \\
E & = & {fL\pi\over 8} \left[{1\over (\beta + \alpha)^2}
+{1 \over (\beta - \alpha)^2}\right], \nn \\
{\rm and}~~S & = & {fL\pi \over 4}[{1\over \beta + \alpha}
+{1 \over \beta - \alpha}].
\label{eq:therfive}
\eea
Since we have massless particles in one spatial dimension, they
can be either right-moving, with $e_r = p_r$, or left-moving, with
$e_r = -p_r$. The corresponding distribution functions then become
\ben
\rho_R  =  {1 \over e^{(\beta + \alpha)e_r}\pm 1}~~~~
{\rm and}~~~~~~~~~
\rho_L  =  {1 \over e^{(\beta - \alpha)e_r} \pm 1}.
\label{eq:thersix}
\een
Thus, the combinations $T_R = 1/(\beta + \alpha)$ and
$T_L = 1/(\beta - \alpha)$
act as effective temperatures
for the right- and left-moving
modes, respectively, resulting in
\ben
{1\over T} = {1\over 2} \left({1\over T_L} + {1\over T_R} \right).
\label{eq:thersixa}
\een
In fact, all the thermodynamic quantities
can be split into left- and  a right-moving pieces:
$E = E_R + E_L~~~~P = P_R + P_L~~~~~S = S_R + S_L$.
The various quantities $E_L, E_R, P_L, P_R, S_L, S_R$
may be read off from Equation~\ref{eq:therfive},
\ben
T_R = {\sqrt{8 E_R \over L\pi f}} = {4 S_R \over \pi f L}
~~~~~~~~~~~~~T_L = {\sqrt{8 E_L \over L\pi f}} = {4 S_L \over \pi f L}.
\label{eq:therseven}
\een

\subsubsection{Near-Extremal Thermodynamics from the Long String}

It is now clear that the statistical mechanics of this free gas in fact
reproduces the thermodynamics of the near-extremal black hole. The
relationships of
Equation~\ref{eq:therseven} become the relations
Equation~\ref{eq:fdseventeen} if
\ben
fL = 8\pi RQ_1Q_5.
\label{eq:thereight}
\een
Given this agreement, it is obvious that the thermodynamic
relations of
Equation~\ref{eq:fdeighteen} are also exactly reproduced.
In particular, the physical temperature $T$ agrees with the
Hawking temperature, as is clear from Equations~\ref{eq:thersixa} and
\ref{eq:fdfifteen}.

Recall that the size of the $x^5$ circle is $2\pi R$. Thus, one way
to satisfy the relation in
Equation~\ref{eq:thereight} is to consider
$L = 2\pi R$ and $f = 4Q_1Q_5$. This would be the situation if
we had $Q_1$ D1-branes individually wrapping the $x^5$ direction.
Another way to satsify Equation~\ref{eq:thereight} is to take
$L = 2\pi Q_1Q_5 R$ and $f = 4$. This is the content for the long
string, where all the D1-branes join up to form a single multiply
wound string.

It might appear that thermodynamics does not
distinguish between
these two configurations. However,
the above thermodynamic relations assume that the system is extensive \cite{maldasusskind}. One
condition for that is that $T_L L, T_R L \gg 1$. It follows from
Equation~\ref{eq:therseven} that this would require
\ben
S_L, S_R \gg f.
\label{eq:thereighta}
\een
There is an important difference between the two cases.
When multiple branes are each singly wound, the energy
of a single particle is $n/R$ for some integer $n$. Thus,
a near-extremal configuration may be obtained from an extremal
configuration by adding $n$ pairs of left- and right-handed quanta,
leading to an excitation energy $\Delta E = 2n/R$. For
the long string, similarly, we would have $\Delta E = 2n/Q_1Q_5 R$, since the length is now $2\pi Q_1Q_5R$. From this it is
apparent that for the multiple singly wound branes,
\ben
S_L = 2\pi{\sqrt{Q_1Q_5(N +n)}}~~{\rm and}~~S_R = 2\pi{\sqrt{Q_1Q_5n}},
\label{eq:thernine}
\een
whereas for the multiply wound long string
\ben
S_L = 2\pi{\sqrt{Q_1Q_5N +n}}~~{\rm and}~~S_R = 2\pi{\sqrt{n}}.
\label{eq:therninea}
\een
It is then clear that the extensivity condition (Equation~\ref{eq:thereighta})
is always satisfied for the long string, whereas it is violated
by the singly wound branes when $Q_1,Q_5$ and $N$ are all comparable
and large.

The multiply wound string has much lower energy excitations,
with a
minimum gap of $1/Q_1Q_5R$. This is consistent with the
conditions for the validity of statistical mechanics for black
holes \cite{preskilletal}. The point is that for statistical mechanics
to hold, the temperature of a system must be such that the specific
heat is larger than one.
It follows from the relations in
Equation~\ref{eq:fdeighteen} that this requires (up to numerical factors of
order one)
\ben
\Delta E > {1\over RQ_1Q_5},
\een
which indicates that the system should have an energy gap
$1/Q_1Q_5R$. This is exactly what the long string provides.

\subsection{A More Rigorous Treatment for Extremal Black Holes}

In the above discussion we adopted a simple model of the D1-D5
bound state and used it to compute the entropy of momentum
excitations. We now discuss more rigorous results on the count of
states that can be obtained by using the fact that we are counting
supersymmetric states when we compute the entropy of an extremal black
hole.  In fact, such a chain of arguments was used in the first exact
calculation of the microscopic entropy of a black hole
by Strominger \& Vafa \cite{stromingervafa}.
We outline the main ideas of such a treatment here. (This subsection
assumes some familiarity with dualities in string theory.)

\bigskip

(A):\quad
 We count not the actual states of the system but an
``index.'' The simplest example of an index is the Witten index of a
supersymmetric system ${\rm Tr}[(-1)^Fe^{-\beta H}]$ ($F$ is the
fermion number and $H$ is the hamiltonian).  This  index counts the
number of bosonic ground states minus the number of fermionic
ground states. The ground states are maximally supersymmetric
states, so if one were interested in states preserving maximal
supersymmetry, then the Witten index would provide useful
information. Note, however, that the index can only give a lower bound
on the number of supersymmetric states, since some states are
counted with a negative sign. As we change the parameters of the
theory, states can appear and disappear in Bose-Fermi pairs, but the
index is a robust quantity.

In the black hole context, we
are
interested not in the ground
states of the system but in states preserving some fraction of the
supersymmetry. (In fact, the states we want carry momentum and thus a
nonzero total energy.) We need a generalization of the Witten
index that will carry nontrivial information about the count
of excited states of the
system.  Of course, we will again obtain a lower bound to the state
count for the given quantum numbers; we then hope that the bound will
approximate the actual state count well
enough to permit a meaningful comparison to
the Bekenstein entropy.

\bigskip

(B):\quad
Let us consider the black hole discussed by Strominger \& Vafa~\cite{stromingervafa}.
Take
Type IIB theory but let the complete 10-dimensional spacetime be $K3\times
S^1\times M^5$, where
$M^5$ are the noncompact directions, $K3$ is a special kind of
compact 4-dimensional surface, and
$K3$
along  with the circle $S^1$ makes up the
five compact directions. Since
one compact direction is $S^1$, there is really no difference between
Type IIA and Type IIB theories here; we can go from one to the other
by a T-duality.

What kind of charges can we place on the compact directions to make
the black hole? In analogy to the  discussion above of the black hole
in $T^4\times
S^1\times M^5$, we can wrap various branes around the cycles of the
compact directions, making sure that they all stretch along the
$S^1$. Then we can put a
momentum charge along the $S^1$ direction and obtain a black hole
carrying a momentum charge $N_p$ along with the charges arising from
the wrapped branes.

We need some way to describe the charges carried by the branes. Type
IIB theory has
1-, 3-, and 5-dimensional branes.
On the other hand, $K3$ has
0-, 2-, and 4-dimensional cycles
on which something can be wrapped, so we can
wrap all
these branes on the $K3\times S^1$ space as desired.  As mentioned
above,  we can
view the theory as Type IIA on $K3\times S^1$, and then we note that
there is a duality map that relates Type IIA compactified on $K3$ to
heterotic
string theory  compactified on $T^4$.
Type IIA compactified on $K3 \times S^1$ maps to heterotic string theory compactified on $T^5$,
and here the charges are easy to understand. The
charges that are dual to
the wrapped branes in
Type
IIA now arise in the heterotic theory from
momentum and winding of the  elementary heterotic string on the
compact directions of
the theory. The possible charge states are characterized by points on a lattice
$\Gamma^{21,5}$, which means a lattice with 21 positive signature
directions and 5 negative signature directions.
The only relevant property of these charges
is that a bilinear invariant, which we may call
$Q_F^2$, gives the invariant length squared of a charge vector
in this space.  This invariant is a generalization of $n_pn_w$, the
product of winding and
momentum charges in any one direction; such a product is invariant
under T-duality, since the two kinds of charges get interchanged.

   \bigskip

(C):\quad
Since
we will be looking for a quantity that will not vary with the
continuous parameters of the system, we can take the size of the
$S^1$ to be much larger than the
size of the $K3$.  Then the physics of the wrapped branes looks like
a $(1+1)$-dimensional sigma model, with the space direction being the
$S^1$ cycle.
It is possible to count the degeneracy of the ground states of the wrapped
branes and infer that the target space of this sigma model must be
essentially the symmetric
product of $Q_F^2/2+1$ copies of $K3$  ($Q_F^2$ as defined is
an even number). Here the term ``essentially'' means that this target
space may be a
deformation of the stated one,
but the topological properties of the
space would remain unchanged, and the space would remain hyper-K\"ahler,
which means
that the number of supersymmetries would also remain the same. The
$Q_F^2/2+1$ copies of $K3$ are analogous to the $Q_1Q_5$
strands of the string that we
took as the vibrating elements in the heuristic treatment of Section~5.5 above;  each point on each strand could move on $T^4$ in that
case, and the $T^4$ is now
replaced by $K3$. The fact that we take a symmetric product amounts
to saying
that the strands are identical and must be so treated in
any quantum mechanical count.

We now consider $n_p$ units of momentum along the $S^1$ direction of
this sigma model, just as we did in the case of $T^4$
compactification. If we hold $n_p$
fixed and take the size of $S^1$ to infinity, then we will be looking
at modes of very long wavelength, longer than any length scale that
existed in the sigma model.
The physics of such modes must be a conformal field theory  (CFT) in
$1+1$ dimensions, and in particular we can separate the excitations
into left- and right-moving
modes. If we do not excite the right movers and let the left movers
carry the given momentum, then we will have states with half the
maximal
supersymmetry: The right supercharges will annihilate the state
and the left ones will not.

If we have such a conformal theory with $N=2$ or larger
supersymmetry, then it can be shown that the elliptic genus
$${\rm Tr} [(-1)^{J_L-J_R}e^{-\beta H}y^{J_L}]$$
   is also a robust quantity that does not change when the moduli are
varied. Here $J_L$ and $J_R$
are integer-valued $U(1)$ charges for the left
and right sectors; these
are the $U(1)$ symmetries that rotate the two supercharges of $N=2$
supersymmetry into each other. We want only the count of states when
the charges are
large, since only then can we compare
the count
with the Bekenstein entropy
computed from the classical gravity solution. In a CFT, the degeneracy
of states at an
excitation level  $n$ above the ground state (in, say, the left sector)
is given in terms of the central charge of the CFT by the formula
$$d(n,c)\sim e^{2\pi{\displaystyle \sqrt{nc\over 6}}}.$$
The central charge for a supersymmetric sigma model with a
hyper-K\"ahler space is easy to calculate. The dimension of the target
space must be of the form 4$k$.
Each bosonic direction contributes 1 to the central charge, and its
corresponding fermion gives an extra 1/2.   Thus $c=6k$.
Putting in the values
 of $k$ and $n=n_p$, we get
$$S=\ln d(n_p, Q_F)\sim 2\pi \sqrt{n_p \left({Q_F^2\over 2}+1 \right)},$$
which agrees with the Bekenstein entropy for the classical geometry
with the same charges.

\bigskip

(D):\quad
Although the above expression was derived for large $n_p$,
it is also possible to
compute the elliptic genus for the symmetric product of copies of
$K3$  without making
any approximation. One can relate the elliptic genus when the target
space is a  symmetric product of $k$ copies of the space $X$ to the
elliptic genus when there
is only one copy of the space \cite{dmvv}. The derivation of this
relation is combinatorial; one symmetrizes over bosonic states and
antisymmetrizes over
fermionic states. But the case where the target space is only one
copy of $X$ is  the theory of a single string on $X$, which is a
simple CFT to handle, and the
elliptic genus can be explicitly written.

If we apply the above method to the case where the compact space is
$T^4$ instead of $K3$, we find that the elliptic genus is zero. This
happens because there are
fermion zero modes on the torus, which relate a set of bosonic states
to an identical spectrum of fermionic states, and there is
a cancellation in the elliptic genus
and we get no useful information about the state count.
Maldecena et~al showed \cite{tfour} that one can use the fact that we have $N=4$ supersymmetry  in
the CFT to argue
that the quantity
$${\rm Tr} [(-1)^{J_L-J_R}J_R^2e^{-\beta H}y^{J_L}]$$
is a topological invariant in this case. This quantity is nonzero,
and it turns out to give for large charges a state count that reproduces
the Bekenstein entropy.

\subsection{The Gauge Theory Picture}

We now show how the above picture of a nonlinear sigma model
follows from gauge theory considerations. As explained above, we are
interested in the long-distance limit of the theory of the D1-D5
system. This system has $(1,1)$, $(5,5)$, and $(1,5)$ strings and has
$(4,4)$ supersymmetry when viewed as a 1-dimensional theory living
along $x^5$.  There is a global R symmetry
which rotates the various
supercharges: $SO(4) = SU(2)_L \times SU(2)_R$. In fact, this R
symmetry is nothing but the rotation symmetry in the four transverse
directions.  The lowest oscillator modes of the open strings lead to
a $U(Q_1) \times U(Q_5)$ gauge theory with
the following supersymmetry multiplets:
\begin{enumerate}
\item The states of the $(1,1)$
and $(5,5)$ strings along the transverse directions $x^1 \cdots x^4$
form a ``vector multiplet.'' There are thus four scalars transforming
as $(2,2)$ of the R symmetry group and four fermions. The left-handed
fermions transform as $(1,2)$ of R symmetry, and the right-handed
fermions transform as $(2,1)$.
\item The states of the $(1,1)$ and
$(5,5)$ strings with polarizations along the $T^4$ directions form a
``hypermultiplet.'' This also has four scalars transforming trivially
under the R-symmetry group and has four fermions. The left-moving
fermions now transform as $(2,1)$ and the right-moving fermions as
$(1,2)$.
\item The states of the $(1,5)$ and $(5,1)$ strings lead to
hypermultiplets whose transformation properties under R symmetry are
the same as those of
the previous hypermultiplets.
\end {enumerate}

We want the branes to be on top of each other. This requires the
vector multiplet scalars to have vanishing expectation values, since
these represent the relative distances between individual branes
in the transverse space. On
the other hand, the scalars in the hypermultiplet will have expectation
values. This is known as the Higgs branch of the theory. As a
result, the vector multiplets typically acquire a mass.\footnote{A $U(1)$ component of the vector multiplet remains massless,
representing the overall motion of the branes in transverse
space. However, the $U(1)$ parts of the fields are decoupled from the
$SU(Q_1) \times SU(Q_5)$ part and can be ignored in this discussion.}
The hypermultiplet scalars are massless. However, they are not all
independent of each other but are related by a set of conditions.
It is necessary to solve for the independent set of fields and find the
low-energy theory of these fields.
Even at the classical level, this is a complicated
affair. It has been pursued elsewhere \cite{maldathesis,hassanwadia}, although
a detailed low-energy theory has not yet been rigorously derived.

Nevertheless, it is possible to obtain an important universal quantity
of the low-energy conformally invariant theory---the central
charge.  This is because the central charge
is related, by superconformal symmetry,
to the anomaly of the R-symmetry current. The R-symmetry current is
anomalous because the left- and right-moving fermions behave
differently under R-symmetry transformations. At high energies
where the theory does not have superconformal invariance but does have
$(4,4)$ supersymmetry, the
anomaly coefficient $k$ may be easily computed in terms of the number
of hypermultiplets and vector multiplets, and it turns out to be $k =
Q_1Q_5$. Because of the well-known 't Hooft anomaly matching
conditions, this coefficient is the same at low energies, where the
theory flows to a superconformal theory. Superconformal symmetry then
determines the central charge of the superconformal algebra in terms
of the anomaly coefficient $c = 6 Q_1 Q_5$.
The asymptotic density of states in a conformal field theory can be
now determined in terms of the central charge, and for BPS states this
leads to the expression for entropy given above.

There is a slightly different but equivalent approach to the problem.
D1-branes in the D5-branes can be considered as ``instanton strings''
of the 6-dimensional $U(Q_5)$ gauge theory on the D5-brane
\cite{douglasbwb}. Actually
these are not really instantons but rather solitonic objects.
Consider configurations of the gauge theory that are
independent of $x^5$ and time $x^0$. Such configurations may be
regarded as configurations of a Euclidean 4-dimensional
gauge theory living
in the $x^6 \cdots x^9$ directions. Instantons
are such configurations, which are self-dual in this 4-dimensional
sense. From the point of view of the 6-dimensional theory, they are
thus solitonic strings. The number of D1-branes is equal to the
instanton number $Q_1$. These instantons form a continuous family of
solutions with the same energy---the corresponding parameters form
the ``instanton moduli space.'' The low-energy excitations of the
system are described by the collective excitations or dynamics on
this moduli space. Because of translation invariance of the original
configurations along $x^5$ and $x^0$, the collective coordinates are
functions of $x^5$ and $x^0$---their dynamics is thus given by a
$(1+1)$-dimensional field theory that is
essentially a sigma model
with the target space as the moduli space of these instantons. This is
the sigma model discussed in the previous section. In fact, the
independent moduli are obtained by solving a set of equations
identical to the equations that determine the independent set
of hypermultiplet scalars.

It turns out that in the orbifold limit, the theory of these instanton
strings may be interpreted in terms of a gas of strings
\cite{dmvv,verlindes} that are wound around the $x^5$ and move along
the $T^4$ with a total winding number $k = Q_1Q_5$. These strings
can be singly wound or multiply wound. The ``long string''
used in the
previous sections is nothing but the maximally wound sector. As we have
argued, this is the entropically favored configuration.

In the orbifold limit, however, the
supergravity approximation is not valid, which is why one has
to turn on some of the supergravity moduli fields to go to the correct
regime. In fact, by turning on a NS-NS two-form field,
one can get rid
of the Coulomb branch altogether, forcing the branes to lie on top
of each other, thus ensuring that we are dealing with the Higgs
branch of the theory \cite{seibergwittend1d5,dijkgraf}. The behavior
of the system with such moduli turned on has been further analyzed
elsewhere \cite{martinecwadia}.

\subsection{Other Extremal Black Holes}

There is a wide variety of black hole solutions in string theory
(for reviews and references see e.g.\ Reference~\cite{cveticyoum}), and numerous studies have related them
to states in string theory.
One notable example is a 4-dimensional black hole
with four kinds of charges \cite{fourdholes}, where once again
there is an exact agreement of the extremal entropy. Near-extremal
limits of these black holes have also been studied; there is, in
fact, a natural analog of the long string approximation in which these
black holes are regarded as intersecting branes in 11-dimensional
supergravity \cite{fourdeleven}. Rotating versions of such
5- and 4-dimensional black holes also provide exact microscopic
derivations of the extremal entropy \cite{rotating}.

\section{BLACK HOLE ABSORPTION/DECAY AND D-BRANES}

In classical gravity, black holes absorb matter via standard
wave scattering by an effective potential arising from the
gravitational field.
Consider the absorption of some field. An
incident wave is partly scattered back and partly absorbed, and the
absorption cross section can be calculated by solving the relevant
wave equation in the black hole background. Hawking radiation is
a quantum process. However, the radiation rate can be obtained
by considering a black hole in equilibrium with its environment
of radiated quanta. The principle of detailed balance then relates
the radiation rate to the absorption cross section by the relation
in Equation~\ref{eq:four}.

We are interested in the decay of a slightly nonextremal black hole
by Hawking radiation. The dominant decay mode is the emission
of massless neutral particles. It is clear that the black hole
will evaporate until the excess mass over extremality has been radiated
away,
leaving an extremal black hole that is stable.
Note that for such black holes, the specific heat is positive (in contrast to neutral black holes, which have negative specific heat),
so that this evaporation is a well-defined process. Furthermore, as
long as the extremal limit corresponds to a large black hole, one
would expect that at all stages of the decay process, semiclassical
approximations are valid for these large black holes.

In the microscopic models described above, the processes of absorption
and radiation appear to be rather different. Consider a
closed string incident on the intersecting D-brane configurations
representing black holes. When the closed string hits the D-brane, it
can either reflect back or split up into open strings whose ends are
attached to the brane. The open strings then do not leave the brane
and hence do not re-emerge in the asymptotic region. Thus, there is
a finite probability
that the closed string quantum will be absorbed
by the black hole, which should be calculated using standard rules
of quantum mechanics. Radiation is the reverse of this process: Open
strings can join up to form closed strings, which can then escape from
the brane system. At low energies, the closed string that is propagating
in the bulk of spacetime
may be replaced by the quantum of some supergravity field,
whereas the open strings may be replaced by quanta of the
low-energy gauge theory living
on the brane. We are interested in
this low-energy limit.

\subsection{Classical Absorption and grey-body Factors}

To compare the absorption or radiation rates of semiclassical black
holes and D-brane configurations, we must compute the classical
absorption cross sections of the relevant fields in the black hole
background. We are interested in
the low-energy behavior of these absorption cross sections.

Let ${\cal \phi}$ denote all the supergravity fields
collectively. Then the wave equation satisfied by a test field in the
classical background ${\cal \phi}_0$ can be obtained by substituting
the perturbed field ${\cal\phi} = {\cal\phi}_0 + \delta {\cal\phi}$ in
the supergravity equations of motion and retaining the part linear in
$\delta{\cal\phi}$.  The equations of motion would generically couple
the various perturbations,
leading to a complicated set of
equations.  However, various symmetries of the background classical
solution allow us to isolate special perturbations that satisfy
simple decoupled equations.

For the 5-dimensional black hole we are discussing, there are 20
such bosonic perturbations, called minimal scalars, that satisfy decoupled massless minimally
coupled Klein-Gordon equations. As we
shall see, these have the largest absorption rate and hence dominate
the Hawking radiation.  Examples of such fields are
traceless parts
 of the components of the 10-dimensional metric,
which have polarizations
along the $T^4$ in the $(x^6 \cdots x^8)$ direction.

Scattering of massless particles by 4-dimensional black holes
was studied in the 1970s \cite{unruhpage}, and
the low-energy limit of the s-wave absorption cross section for massless
minimally coupled scalars was found to be exactly
equal to the horizon area for Schwarzschild and charged
Reissner-Nordstrom black holes. A similar calculation for the
near-extremal 5-dimensional black hole described above also showed
that the leading term of the absorption cross section is again the
horizon area \cite{dhar,dasmathur}. It was soon realized, however,
that this is in fact a universal result: For all spherically
symmetric black holes in any number of dimensions, the s-wave
cross section of minimally coupled massless scalars is always the area
of the horizon \cite{dasgibbonsmathur}.

The actual calculation of the cross section involves solving the
wave equation analytically in various regimes and then matching solutions in the overlapping regions. As an example,
we describe the essential steps that led
to the result obtained in  \cite{dasgibbonsmathur}.

Consider a metric of the form
\ben
ds^2 = -f(r) dt^2 + h(r) \left[dr^2 + r^2 d\Omega_p^2 \right],
\label{eq:absone}
\een
where $d\Omega_p^2$ denotes the line element on a unit $p$-sphere.
Let the horizon be at $r = r_H$ so that $f (r_H) = 0$. We want to solve
the minimally coupled massless equation $\nabla^2 \phi = 0$ for
a spherically symmetric scalar field $\phi (r,t)$. Using
time translation invariance, we decompose the field into
fields of definite frequency, $\phi_\omega (r)$, through
$\phi (r,t) = e^{-i\omega t}\phi_\omega (r)$.
The frequency component $\phi_\omega (r)$ satisfies
\ben
[(r^p F(r) \partial_r)^2 + \omega^2 R^{2p} (r)]
\phi_\omega (r) = 0,
\label{eq:abstwo}
\een
where
\ben
F(r) = \{f(r) [h(r)]^{p-1}\}^{1/2}~~~~~~~~~~
R(r) = r [h(r)]^{1/2}.
\label{eq:absthree}
\een
Let $l$ denote the largest length scale in the classical solution.  We
then solve the equation to lowest order in $\lambda =
\omega l$. Then there are two regions in which the equation can be
easily solved. In the outer region $\omega r > \lambda$, the equation
becomes that of a scalar field in flat background and may be solved
in terms of Bessel functions. In the near-horizon region
$\omega r < \lambda$,
the function $R(r) \rightarrow R_H = R(r_H)$ may be replaced and the equation
may be solved after a simple change of variables.
We must then match the two solutions in the overlapping region.

The physical input is a boundary condition. It is necessary to impose a
boundary condition so that there is no outgoing wave at the horizon.
In the asymptotic region, there is both an incoming and an
outgoing wave, and matching the solution to the near-horizon solution yields
the ratio of the outgoing and incoming components and hence the
absorption probability. This can be converted into an absorption
cross section $\sigma (k)$ for a plane wave of momentum ${\vec k}$
using standard techniques. The result is, as advertised,
\ben
{\rm Lim}_{\omega \rightarrow 0}\,\sigma (k) = {2 \pi^{(p+1)/2}\over
\Gamma({p+1 \over 2})} R_H^p = A_H,
\label{eq:absthreea}
\een
where $A_H$ is the horizon area and the last equality follows from
the form of the metric (Equation~\ref{eq:absone}).

For the 5-dimensional black hole, the absorption cross section for
such minimally coupled scalars may be calculated in the so-called
dilute gas regime defined by
\ben
r_0, r_N \ll r_1, r_5~~~~~~{r_1 \over r_5} \sim {r_0 \over r_N}
\sim O(1)
\label{eq:absfour}
\een
when the energy $\omega$ is in the regime
\ben
\omega r_5 \ll 1~~~~~{\omega \over T_L} \sim O(1)
~~~~~~{\omega \over T_R} \sim O(1).
\label{eq:absfive}
\een
The various parameters of the classical solution have been
defined in previous sections.
The final result is
\cite{maldastrominger}
\ben
\sigma (\omega) =
2\pi^2 r_1^2 r_5^2 {\pi \omega \over 2}
{(e^{\omega/T_H} - 1) \over (e^{\omega/2T_L} - 1)(e^{\omega/2T_R} - 1)}.
\label{eq:abssix}
\een
Remarkably, the cross section appears as a combination of thermal
factors, even though it is a result of a solution of the Klein-Gordon
equation. When $T_L \gg T_R$ and $\omega \ll T_L$, one has $T_H \sim
{1\over 2}T_R$ and the cross section becomes exactly equal to the
area of the horizon. The expression (Equation~\ref{eq:abssix}) is called a grey-body factor because of
the nontrivial energy dependence.

Finally, let us briefly discuss the absorption of higher angular
momentum modes and higher spin fields. Writing the relevant
wave equations, we  find for such modes an additional
``centrifugal'' potential near the horizon that suppresses absorption
of these modes by the black hole at low energies
\cite{unruhpage}. The exceptions
 are
modes of spin-1/2 fields that are supersymmetric partners of the
minimal scalars. When these modes have their orbital angular momentum,
the absorption cross section is twice the area of the horizon, in
agreement with expectations from supersymmetry.

\subsection{D-brane Decay}

A first-principles calculation of the decay of a slightly nonextremal
5-dimensional black hole would involve a strongly coupled gauge
theory of the D1-D5 gauge theory coupled to the bulk supergravity
fields.
As discussed above, the gauge theory is
difficult to analyze even in weak coupling. However, we have a fairly
good effective theory---the long string model---and weak coupling
calculations in the model give
exact strong coupling answers.  One
might wonder whether the model may be used to compute decay rates.
Once again one may hope that weak coupling calculations could give
reasonable answers for the same reasons that they succeeded in
predicting the right thermodynamics.  In this subsection, we
show that this is indeed unambiguously possible for a certain set of modes,
and indeed weak coupling calculations for these modes agree
exactly with the semiclassical results described in the previous
subsection.

The theory of the long string is a $(1+1)$-dimensional massless
supersymmetric field theory with four flavors of bosons $\phi^I$, the
index $I$ referring to the four directions of the $T^4$ along
$x^6 \cdots x^9$ and their corresponding fermionic partners. We want
to figure out how these degrees of freedom couple to the supergravity
modes in the bulk. In fact, we are mostly interested in
supergravity modes that are minimal scalars in the dimensionally
reduced theory. Out of these 20 scalars, it is particularly easy to
write down the interaction of the long string with the transverse
traceless components of the 10-dimensional graviton $h_{IJ}$,
with indices $I,J$ lying
along the $T^4$. At low energy, the relevant action is
\ben
{\cal T}\int d^2\xi \partial_\alpha \phi^I \partial^\alpha \phi^J
g_{IJ},
\label{eq:absseven}
\een
where the $\xi^\alpha$ denote
 the coordinates on the long string
world sheet and $g_{IJ}$ is the metric on the $T^4$. ${\cal T}$ is a
constant.  This form of the action follows from the principle of
equivalence.  When there is no supergravity background, the action
should be a free action given by
Equation~\ref{eq:absseven} with $g_{IJ}$
replaced by $\delta_{IJ}$.  The fields $\phi^I$ transform as vectors
under the local $SO(4)$ transformations of the tangent space of the
$T^4$. This requires that at low energies the action in the presence
of a background is given by Equation~\ref{eq:absseven}.

If we expand the metric as
\ben
g_{IJ} = \delta_{IJ} + {\sqrt 2}\kappa_{10} h_{IJ}(\xi,\phi^I,
x^1\cdots x^4)
\label{eq:absnine}
\een
we immediately get the coupling of $h_{IJ}$ with the long string degrees
of freedom. In Equation~\ref{eq:absnine}, we have made explicit the dependence
of the supergravity field on the coordinates
on the long string world sheet $\xi = (x^0,x^5)$, the $T^4$ coordinates
$\phi^I$ (which are the fields of the long string theory), and the
coordinates transverse to the black hole $(x^1 \cdots x^4)$.
The 10-dimensional
gravitational coupling, $\kappa_{10}$, is related to the string coupling and
string length by
\ben
\kappa_{10}^2 = 64 \pi^7 g^2 l_s^8.
\label{eq:absninea}
\een
The factor
${\sqrt 2}\kappa_{10}$ in Equation~\ref{eq:absnine} is determined by requiring that
the field $h_{IJ}$ be canonically coupled and follows from the bulk
supergravity action (Equation~\ref{eq:hsixa}). There will be other terms in
the action, but these would contain more derivatives and are suppressed
at low energies.


It is clear from Equation~\ref{eq:absseven} that the decay of a nonextremal
state into the mode $h_{IJ}$ is dominated by the process of
annihilation of a pair of modes of vibration from the long string, one of which is left-moving and the other right-moving (see Figure~3). Because we are working in the limit
where the radius $R$ of the $x^5$ circle is much larger than the size
of the $T^4$, we can ignore the dependence of $h_{IJ}$ on
$\phi^I$. Furthermore, we are interested in computing s-wave
absorption/decay, which depends
on the transverse coordinates
only through the radial variable.

\begin{figure}
\epsfysize=6cm \epsfbox{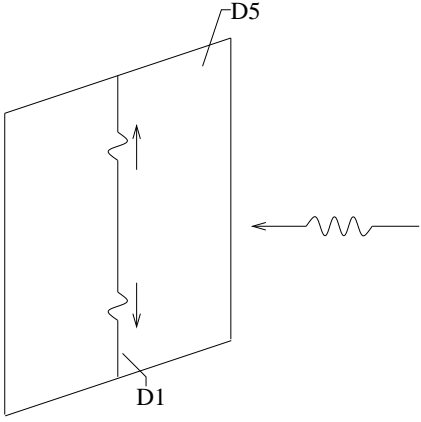}
\caption{Absorption of supergravity mode by D-brane.}
\end{figure}

The constant ${\cal T}$ is an effective tension of the long string and
depends
on the details of the rather complicated D1-D5 bound
state. It appears that the action we have written has little
predictive power unless we figure out these details. However, it
has been recognized \cite{dasmathur} that the cross section in question does
not depend on such details. This is because the free kinetic term of
the fields $\phi^I$ has to be canonically normalized. Because the
interaction is also quadratic in $\phi^I$, normalization thus completely removes
the constant ${\cal T}$ from the action. The amplitudes that
result from this action are therefore universal. In
the following, we give the basic steps for calculation of the decay
rate when the mode $h_{IJ}$ carries no momentum along $x^5$ and is in
an s-wave
in the transverse space. This corresponds to the decay into
neutral scalars from the noncompact 5-dimensional point
of view.

Consider the annihilation of two long string modes: ($a$) one with
polarization $I$ and momenta $p = (p^0, p^5)$ [along the
$(x^0,x^5)$
direction] and ($b$) one with polarization $J$ and momenta $q= (q^0,q^5)$
into an $h_{IJ}$ bulk mode with momenta $k = (k^0,k^1, \cdots, k^4,0)$
along the
$(x^0,\cdots, x^5)$ directions. Using Equations~\ref{eq:absseven} and
\ref{eq:absnine}, we find that the decay rate for this process is
\ben
\Gamma(p,q;k) = (2\pi)^2 L \delta(p_0+q_0-k_0) \delta (p_5 + q_5 - k_5)
~{ 2 \kappa^2 (p \cdot q)^2 \over (2p_0 L)(2q_0 L)(2 k_0 V L V_4)}~
{V_4d^4 k \over (2\pi)^4}.
\label{eq:absten}
\een

The delta functions impose energy conservation and momentum conservation
along the $x^5$ direction (momentum is not conserved in the other
directions because of Dirichlet
conditions). The denominators come
from normalization of the modes: the open string modes are
normalized on the circle along $x^5$ of circumference $L = 2\pi Q_1 Q_5 R$
whereas the closed string field is normalized in the entire space with
volume $V L V_4$, where $V_4$ is the volume of the noncompact four
spatial dimensions $x^1 \cdots x^4$.

To obtain the total cross section for production of a scalar, one must now average over all initial states. Because these states are drawn
from a thermal ensemble,
the decay rate is
\ben
\Gamma(k) = ({L \over 2\pi})^2\int_{-\infty}^\infty dp_5
\int_{-\infty}^\infty dq_5~\rho(p_0,p_5) \rho(q_0,q_5) \Gamma(p,q;k),
\label{eq:abseleven}
\een
where $\rho(p_0,p_5)$ denotes the Bose distribution functions discussed
in the previous section. The integral may be evaluated easily and
the answer is
\ben
\Gamma(k) = 2 \pi^2 r_1^2 r_5^2 {\pi \omega \over 2}
{1 \over
(e^{{\omega \over 2 T_L}} - 1)(e^{{\omega \over 2 T_R}} - 1)}.
\label{eq:abstwelve}
\een
Finally, the decay rate is converted into an absorption cross section by
multiplication by an inverse Bose distribution of the supergravity
mode:
\ben
\sigma(\omega)^{{\mathrm D-brane}} = 2 \pi^2 r_1^2 r_5^2 {\pi \omega \over 2}
{e^{{\omega \over T}} - 1 \over
(e^{{\omega \over T_L}} - 1)(e^{{\omega \over T_R}} - 1)}.
\label{eq:absfourteen}
\een
We have expressed the answer in terms of the parameters in the
classical solution by using the length of the effective string,
\ben
L = 2\pi Q_1 Q_5 R = {8\pi^4 r_1^2 r_5^2 V R \over \kappa^2}.
\label{eq:absthirteen}
\een
The physical temperature $T$ is related to $T_L$ and $T_R$ by
Equation~\ref{eq:thersixa}. Thus, when $T_L \gg T_R$, we have $T \sim 2T_R$.
However, we know that the quantities $T_L$ and $T_R$ above are in fact
exactly the same as the semiclassical quantities and $T = T_H$.
For very low energies $\omega \ll T_L$, one of the thermal factors
above can be simplified. Using
Equations~\ref{eq:fdfourteen}
through \ref{eq:fdsixteen}, we see that in this limit
\ben
\sigma (\omega) = 2\pi^2 (r_1r_5)^2 T_L = {(r_1r_5)^2 S_L
\over RQ_1Q_5} = A_H.
\label{eq:absfifteen}
\een
We have used
Equation~\ref{eq:fdseventeen} and the fact
that in the regime $T_L \gg T_R$ we have $S_L = S = S_{{\mathrm BH}} =
2\pi A_H/\kappa_5^2$, where $\kappa_5^2 = \kappa_{10}^2/2\pi RV$ and $\kappa_{10}^2$ is given by Equation~\ref{eq:absninea}. We have
also used Equation~\ref{eq:fdfourteen} to express $r_1,r_5$ in terms of $g,l_s,
V,Q_1,Q_5$. This is exactly the low-energy classical result.
The fact that the cross section turns out to be proportional to the
horizon area was already known \cite{callanmalda}. The precise
calculation outlined above was performed in Reference~\cite{dasmathur}.

Even before taking this low-energy limit, the entire result
(Equation~\ref{eq:absfourteen}) is in exact agreement with the semiclassical
grey-body factor
\cite{maldastrominger}.

It is clear from Equation~\ref{eq:abstwelve} that the thermal factor in the
decay rate comes from the thermal factors of the long string
modes. In the strict classical limit, we can have absorption but
no Hawking radiation; this comes about because the temperature
in this limit is zero, so the corresponding thermal factor suppresses
emission completely while absorption is still nonzero \cite{dhar}.

The supergravity mode can also split into a pair of
fermions. However, then the thermal factors that appear in
Equation~\ref{eq:abstwelve} are Fermi-Dirac distributions and these go to
a constant at low energies, rather than diverging as Bose factors do.
As a result, the corresponding cross section is suppressed at low
energies. In a similar way, we may consider emission/absorption of
a fermionic supergravity mode. This would require one left-moving
bosonic mode of the long string with a right-moving fermionic mode.
Following the above calculations, it is apparent that the thermal
factor of the left-moving long string mode
gives the right powers
of  energy to lead to a nonzero low-energy cross section, and the
thermal factor for the right-moving mode provides the thermal
factor of the emitted supergravity fermion---and this is a
Fermi-Dirac factor as expected \cite{callanmalda}.

Classical emission of higher angular momentum modes is suppressed by
centrifugal barriers provided by the gravitational field. In the
microscopic model, such emission is again suppressed, but for a
different reason. Note that the bosonic degrees of freedom of the long
string are all vectors under the internal $SO(4)$ of the $T^4$ but are
scalars under the tangent space $SO(4)$ of the transverse
space. Therefore, these can never collide to give rise to modes with
nonzero angular momentum in transverse space. Fermions on the long
string, however, carry transverse tangent space indices (typically as
a R-symmetry index) since they are dimensional reductions of 10-dimensional fermions. Thus, multifermion processes are responsible for
emission of higher angular momentum modes. However, these can be seen
to be necessarily suppressed, since they involve more than two fermion
fields and therefore correspond to higher-dimension operators on the
long string world sheet. Angular-momentum--mode emission is not completely
understood in the long string picture---although the qualitative aspects
have been understood \cite{angularmomentum}.

Similar calculations have been performed for several other
situations, e.g.\ for absorption/emission of charged scalars from both
5- and 4-dimensional black holes \cite{gubserkleba1,maldastrominger}.

The low-energy effective action used above follows from a
Dirac-Born-Infeld (DBI) action of the long string in the presence of supergravity
backgrounds.  One might wonder whether this DBI action can be used to
predict absorption rates for other supergravity modes. It turns out
that this can be done for several other minimal
scalars, namely
the dilaton and the components of the NS $B$ field
along the $T^4$. There are four other minimal scalars in this
background, the couplings of which
cannot be obtained from the DBI
action. Perhaps more significantly, there is a set of other scalar
fields that do not obey minimal Klein-Gordon equations---e.g.\ the
fluctuations of the volume of the $T^4$. If we trust the DBI action to
obtain the couplings of these fields, we get answers that do not
agree with supergravity calculations \cite{fixedscalars}. These cross sections are
suppressed at low energies and it is probably not surprising that
the simplest low-energy effective theory does not work.

\subsection{Why Does It Work?}

The agreement of the low-energy limit of the absorption
cross section---or, equivalently, decay rate---is strong evidence for the
contention that Hawking radiation is an ordinary, unitary quantum decay
process. This is, however, surprising---like the agreement of
near-extremal thermodynamics---since our weak coupling calculation agreed
with a strong coupling expectation. The
agreement of nontrivial grey-body factors is even more surprising
because now not just a single number but an entire function of energy
is correctly predicted by a weak coupling calculation.

Weak coupling calculations also gave the correct answer for the
extremal entropy,
but that case involves
BPS states and
supersymmetry guarantees the agreement. In nonextremal situations,
supersymmetry is broken. However, for small amounts of nonextremality,
we have small excitations over a supersymmetric background and we may still expect nonrenormalization theorems to
ensure that higher-loop effects vanish at low energies. A complete
understanding of this is still lacking. Explicit one-loop
calculations \cite{dasloop} seem to support this scenario. There is
also some evidence for nonrenormalization theorems of the type
required \cite{maldacena2}.

\section{ABSORPTION BY THREE-BRANES}

The above discussion has addressed extremal solutions with
large horizon areas and their excitations. Extremal solutions with
vanishing horizon areas
 generally have singular horizons, for which
semiclassical results cannot be trusted. There are, however, several
exceptions to this rule.
In this section
we consider one of them---the 3-brane in Type IIB supergravity.

\subsection{Classical Solution and Classical Absorption}

The classical solution for $N$ parallel extremal 3-branes along $(x^7
\cdots x^9)$ is given by the (Einstein-frame) metric
\ben
   ds^2 = [f(r)]^{-1/2}[-dt^2 + d{\vec x}^2]
+ [f(r)]^{1/2}[dr^2 + r^2 d\Omega_5^2],
\label{eq:tbone}
\een
where $r = (\sum_{i=1}^6 (x^i)^2)^{1/2}$ is the radial coordinate in
the transverse space, ${\vec x} = (x^7,\cdots, x^9)$ are the
coordinates on the brane, and $d\Omega_5$ is the measure on a unit
5-sphere.  The harmonic function $f(r)$ is given by
\ben
f(r) = 1 + \left({R \over r} \right)^4~~~~~~~~R^4 = 4\pi g_s N l_s^4.
\label{eq:tbtwo}
\een
The dilaton is a constant and the only other nontrivial field is the
4-form gauge field with a field strength
\ben
F_{0789r} = {4 R^4 \over g_s r^5}[f(r)]^{-2}.
\label{eq:tbthree}
\een
The horizon is at $r = 0$ and the horizon area is indeed zero. However, the
spacetime is completely nonsingular. In the near-horizon or
near-brane region $r \ll R$, the metric (Equation~\ref{eq:tbthree})
becomes, in terms of coordinates $ z = R^2/r$,
\ben
ds^2 = \left({R \over z} \right)^2[-dt^2 + d{\vec x}^2 + dz^2] + R^2 d\Omega_5^2.
\label{eq:tbthreea}
\een
This is the metric of the space AdS$_5 \times S^5$, the first factor
being composed of $(t,{\vec x},z)$ while the sphere $S^5$ has a constant
radius $R$.
The AdS$_5$ has a constant negative curvature $1/R^2$,
whereas the sphere $S^5$ has a constant positive curvature. Thus, near
the horizon the spacetime approaches a ``throat geometry,'' where
the length of the throat is along the AdS$_5$ coordinates and the
cross sections are 5-spheres of constant radii $R$.

From
Equation~\ref{eq:tbtwo} it follows that this curvature is much smaller than
the string scale when $g_s N \gg 1$. This is thus the regime where
semiclassical supergravity is valid.

It is possible to calculate the classical absorption cross section of various
supergravity modes by the extremal 3-brane. For s-wave absorption
of minimally coupled scalars, the leading-order
result is \cite{klebanovthreea}
\ben
\sigma = {1\over 8}\pi^4 R^8 \omega^3 = {1\over 32\pi} (\kappa N)^2 \omega^3,
\label{eq:tbfour}
\een
where we have used Equation~\ref{eq:tbtwo}.
The cross section of course vanishes at $\omega = 0$, since the horizon
area vanishes. Also note that we are dealing with an extremal solution
that has zero temperature, so we have absorption but no Hawking
radiation. It turns out that in this case one can also analytically
calculate the low-energy absorption cross section of such scalars for
arbitrary angular momentum \cite{klebawati}.

\subsection{Absorption in the
Microscopic Model}

The microscopic theory of $N$ 3-branes is a $U(N)$ gauge theory in
$3+1$ dimensions with 16 supersymmetries, or $N=4$ supersymmetry.
This is a well-studied theory and is known to be conformally invariant
with zero beta function. It is the dimensional reduction of $N=1$
supersymmetric Yang-Mills in ten dimensions. As we will see below,
a  natural symmetry argument determines
the coupling of
some of the supergravity modes to the brane degrees of
freedom. However, if we consider modes like
the dilaton and
longitudinal traceless parts of the 10-dimensional graviton, the
coupling can be once again read off from
the
equivalence principle. For
example, the longitudinal graviton has to couple to the energy-momentum
tensor of the Yang-Mills theory.  Thus, one can perturbatively
calculate the absorption cross section of such modes (which are
minimal scalars from the 7-dimensional point of view) along the lines
of the calculation in the 5-dimensional black hole.
Again, an exact agreement was found for s-waves \cite{klebanovthreea}, and the agreement was found to extend to all angular momenta \cite{klebawati}.

In this case, the reason for the agreement of the weak coupling
calculation with supergravity answers is much better understood. In
particular, it is
known that the two-point function of the energy-momentum tensor of $N=4$ Yang-Mills theory is not renormalized, so
that the lowest-order result is exact \cite{klebaschwinger}. The imaginary part of the
two-point function in Euclidean space is in turn related to the absorption
cross section, which is thus also protected from higher-order corrections.
There are similar nonrenormalization theorems known for operators coupling
to other modes as well.

\subsection{Nonextremal Thermodynamics}

The extremal 3-brane of supergravity described above is of course
a limit of a general nonextremal solution with finite horizon area
and finite temperature. It is thus natural to expect that the
microscopic description of this is in terms of a Yang-Mills theory at
finite temperature. The thermodynamics of
this Yang-Mills theory was calculated \cite{gubserpeetklebanov} at lowest order in the coupling
constant (i.e.\ a free gas), and it was found that although the dependence
of the thermodynamic quantities on the energy, volume, and $N$ is the
same as in semiclassical thermodynamics of the corresponding black
brane, the precise coefficient differs by a factor of $4/3$.  We know
of no reason why the thermodynamic quantities are not renormalized as
we go from weak to strong coupling, so the discrepancy is not
unexpected.  However, it is significant that the dependences agree---although
the dependence on the energy and the volume is dictated by
scaling arguments (this is a conformally invariant theory), the
dependence on $N$ is not. At weak coupling, we know that there are
$N^2$ degrees of freedom; there is no known argument
why this should
continue at strong coupling.

\section{AdS/CFT CORRESPONDENCE AND HOLOGRAPHY}
%
%
We have presented
a unitary description of black hole
evaporation in terms of the emission of closed strings from the
branes.
This description was at weak coupling; in order to
describe absorption by the black hole of quanta with arbitrary
energies, one must imagine that it is continued to strong coupling.

Let us see what form this continuation to higher energies might take.
In the lowest-order calculation, there was only one string
interaction, where two open strings joined up to the emitted closed
string. At higher orders in the coupling, we expect the emitted string
to interact several times with the branes before escaping to infinity
as radiation.  Let us consider for simplicity the case of the
D3-branes described above. Each such interaction gives one ``hole'' on
the string world sheet, with the boundary of the hole being
constrained to the branes.

From the viewpoint of the dynamics of open strings on the 3-branes,
each hole looks like an open string loop. Thus, as we increase the
coupling, we encounter higher loop processes among the open strings,
in the process of emitting a closed string.

On the other hand, a loop of open
strings can be interpreted as a propagation of a closed string at tree
level (see Figure~4). Thus, these open string multiloop processes might be
reinterpreted as closed string exchanges between the departing quantum
and the branes. But among such closed string exchange must be the
effect of the gravitational effect of the branes, which, crudely
speaking, would attempt to retard the outgoing quantum, redshifting it
to lower energies.

\begin{figure}
\epsfysize=3cm \epsfbox{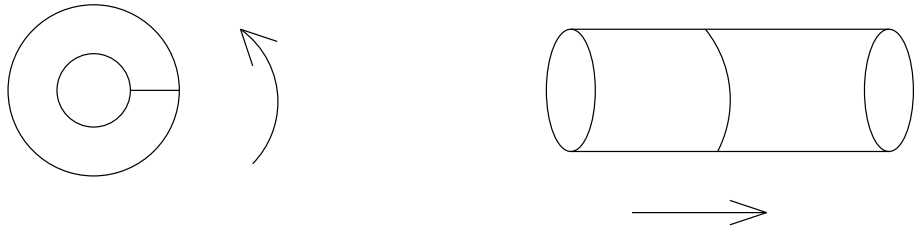}
\caption{Open string loop as a closed string exchange. The arrows
indicate the direction of time in each way of looking at the diagram.
}
\end{figure}

Such a redshift is easily seen if we consider the propagation
of the outgoing quantum in the metric produced by the branes. But now
we ask: As we increase the coupling in the
microscopic calculation, should we take into account the complicated
loop processes of the gauge theory (which are significant when the
open string $g_o$ coupling is not small), as well as the fact that the
metric around the branes will cease to be flat? (This curvature is
significant when the closed string coupling $g_c\sim g_o^2$ ceases to
be small.)
The discussion above indicates that taking both of these effects into account might amount to overcounting.
The effect of open string loops might well be counted in the propagation
of closed string modes (which include gravitons that can be said to
condense to generate the curvature).

It is hard to answer this question from a direct consideration of
loop expansions, since in the domain of large coupling we are unlikely
to have any good expansion
in the number of loops. Thus, the above
considerations are somewhat heuristic. The situation is reminiscent,
though, of the appearance of duality in tree-level string scattering.
The $t$-channel exchange has a sequence of poles, as does the $s$-channel
scattering, and we confront the issue of whether these are two
different effects to be added together in some way or just
dual manifestations of a common underlying dynamics.  In the case of
this string scattering, of course, it is now well understood that the
$s$-channel poles result from an infinite sum over $t$-channel poles and
are not to be considered as additional contributions to the $S$-matrix.

\subsection{The Maldacena Conjecture}

In the much more complicated case of the black hole evaporation
process, Maldacena's bold postulate addressed the
relation between microscopic theory of the branes and the effects of
gravity in 10-dimensional
spacetime.  Maldacena considered the following limit.

Place together a large number $N\gg1$ of D3-branes. Let the string
coupling be weak ($g\ll1$), but
take the limits of large $N$ and small $g$ in such a way that $gN$
tends to a finite value.  Now consider energies of excitation that
are low, so that on the D3-branes we get open strings but only in their lowest states. The
dynamics of these open strings is, as we saw above, given by a
(${\cal N}=4$) supersymmetric
Yang-Mills theory. The quantity $gN\sim g_{{\mathrm YM}}^2N$ is the 't Hooft
coupling of the gauge theory.  At small 't Hooft coupling, we can
study the large-$N$ gauge
theory perturbatively; at large 't Hooft coupling $g_{{\mathrm YM}}^2N\gg1$
such a perturbative  analysis is not possible.

But for $gN\gg1$ we are in a regime where the D3-branes produce an
appreciable gravitational field around themselves, and in fact
distort the spacetime into
the form of
Equation~\ref{eq:tbone}. In particular, as shown above, as we
approach the horizon the geometry becomes the throat geometry
AdS$_5 \times S^5$ given by Equation~\ref{eq:tbthreea}.

Maldacena postulated that we can consider either ($a$)
the gauge theory on the branes (which is given by the open string
interactions at a perturbative level) or ($b$) no
branes, but the space
AdS$_5\times S^5$ and closed string theory (which has gravity as a
low-energy limit) on this curved manifold. These two theories were
dual descriptions of exactly the same underlying theory.

If the gauge theory of the branes and gravity on the 10-dimensional spacetime
are indeed exactly
dual descriptions of the same theory, then
we must have a way of computing
the same quantity in two dual
ways, and thus to see manifestly the consequences of this
duality. To do this, we need an operational definition of the dual map,
which was provided by Gubser et al
\cite{gkp} and Witten \cite{witads}.
Let us rotate the signature of the spacetime to a Euclidean one.  In
this case, the space spanned by ${\vec x}, r$, which is now Euclidean
AdS$_5$, has $r=0$
as a regular point in the interior of the space. Thus, there is no
singularity that might represent the place where the D3-branes were
placed in a gauge
theory description. But this space has a boundary at
$r\rightarrow\infty$, which has the topology of $S^4$. The conjecture
now states that string theory on
this smooth AdS$_5\times S^5$ is dual to ${\cal N}=4$ supersymmetric
$SU(N)$ Yang-Mills on the 4-dimensional boundary of the AdS
space.

Note that in this formulation of the conjecture, all reference to D3-branes has  disappeared. In fact, a geometry  that is AdS$_5\times
S^5$ everywhere
(and not just at small $r$) is an exact solution of string theory
without any branes present. We take the 5-form field strength
$F$ present in Type IIB string
theory to have a constant value with integral
$N$ on the
$S^5$; since this field must be self-dual, it also has a constant
value on the AdS$_5$. The energy density of this field yields the
equal and opposite curvatures of the
$S^5$ and the AdS$_5$ spaces; the radius of curvature in each case is
$R\sim (gN)^{1/4}l_s$ ($l_s$ is the string length). If we had taken a
collection
of D3-branes as above, then the flux of
$F$ on the sphere surrounding the branes would have been
$N$, but only the near-brane geometry would have resembled
AdS$_5\times  S^5$---at large $r$ the space becomes flat. Thus, it
seems better to formulate the
duality without reference to any D-branes, although, as we
discuss below, this makes the issues related to black hole information loss
somewhat more difficult to access.

An important fact that led Maldacena to the duality conjecture was
the agreement of global symmetries between the gravity theory and the
gauge theory.
AdS$_5$ is a maximally symmetric space with a 15-parameter isometry
group. The gauge theory is in four dimensions, so one may expect a
10-dimensional set
of symmetries, which are the translations and rotations on $S^4$. But
the ${\cal N}=4$ supersymmetric $SU(N)$ Yang-Mills theory is a conformal theory,
and the conformal group in four dimensions is indeed isomorphic to the
15-parameter group of isometries of AdS$_5$ space.  Further, the
${\cal N}=4$ gauge theory has four supercharges forming, having an
$SU(4)$ symmetry of rotations among themselves. The gravity theory
has an internal space that is
$S^5$. Thus, when we decompose fields in harmonics on this space in
the process of Kaluza-Klein reduction, the multiplets obtained fall
into representations of
$SO(6)$. But $SO(6)\approx SU(4)$, so both the theories have the same
R-symmetry group.  Further, because of  conformal invariance, the
gauge theory
has 16 conformal supercharges in addition to its 16 regular
supercharges; this agrees with the 32 supercharges of the gravity
theory on AdS$_5\times S^5$.

\subsection{Calculations Using the Conjecture}

Let us return to the issue of how to compare quantities in the two
dual theories.  Let $\Omega$ be the AdS space, and
$\partial\Omega$  its boundary.   Let
us consider for simplicity a scalar field, the dilaton
$\phi$, present in the string theory. If we specify the value
$\phi_b$ of $\phi$ on the boundary $\partial\Omega$,
then we can solve for
$\phi$ in the interior of $\Omega$ and compute the action $S_{{\mathrm SUGRA}}$
of supergravity on this solution. The field
$\phi$, on the other hand, will be dual to an operator in the
Yang-Mills theory, and in this case symmetries fix this
operator to be $Tr F^2$. The duality relation is
\ben
e^{-S^\Omega_{{\mathrm SUGRA}}[\phi_b]}~=~{\int_{\partial\Omega }DA~
e^{-S_{{\mathrm YM}}+\int d\xi\phi_b(\xi)Tr
F^2(\xi)}\over
\int_{\partial\Omega} DA~e^{-S_{{\mathrm YM}}}}.
\label{eq:adsmain}
\een
  Here $\int DA$ represents the
path integral over the gauge theory
variables. The action on the left-hand side is a string action in
general, but at the leading order it can be replaced by
the supergravity action.

    Using the above relation, it is possible in principle to compute $n$-point
correlation functions from supergravity
and compare the results to calculations in the gauge theory.  The
gravity theory is simple (perturbative classical
supergravity to lowest order) when $N$ is large (loops of gravity are
suppressed by $1/N^2$) and  $gN$ is large (because then the
AdS space has a large radius, and stringy corrections, which depend
on $l_s/R$,  can be ignored to leading order).  Thus it is difficult
to compare results on
the gravity side with explicit computations in the Yang-Mills theory,
which has a well-known treatment of the large-$N$ limit but no simple
way to handle a
regime with $g_{YM}^2N\sim gN\gg1$. However, supersymmetry protects the
values of some quantities from changing when the coupling is varied,
and here
agreement is indeed found. Thus, anomalous dimensions of chiral
operators \cite{gkp,witads} and three-point functions of chiral
operators \cite{freed} are found to have the same value when
computed from the gauge theory or from the gravity dual
using Equation~\ref{eq:adsmain}. For
quantities that are not protected in this fashion, the gravity
calculation gives a prediction
for the strongly coupled gauge theory. Examples of such predictions
are the potential between external quarks placed in the gauge theory
\cite{maldastr}, four-point correlation
functions,
and anomalous dimensions of composite operators (which can
be deduced from the four-point functions) \cite{fourpoint}.  If one
considers the gauge theory at a finite
temperature by compactifying the ``time'' direction, then one obtains a
nonsupersymmetric theory with a discrete spectrum of
glueball states \cite{wittemp}.
In this case, the strong coupling calculation has no direct physical
meaning, and only a  weak coupling calculation would give the true
glueball masses (the
coupling flows with scale after supersymmetry is broken;  with
supersymmetry intact, it does not, and each value of the coupling
gives a different well-defined
theory). Nevertheless, there is a surprising  agreement  of the
qualitative features of the glueball spectrum between the gravity
calculation and  lattice
simulations \cite{ooguri}.

\subsection{Holography and the Bekenstein Entropy Bound}

The AdS/CFT correspondence has been studied extensively over the past
two
years [see \cite{adsreview} for review].
This section explains how this
correspondence illustrates another important property of holography
that is in fact its very essence: the Bekenstein bound
\cite{bekenholo}. A general argument states that in any
theory of gravity, the total entropy $S$ of anything in a large
box of volume $V$ and surface area $A$ is bounded by
\ben
S \leq {A \over G},
\label{eq:holone}
\een
where $G$ is Newton's constant. Numerical constants are ignored in
this relation. The form of this bound appears surprising, since in the absence of gravity
we know that entropy is an extensive quantity and should scale as $V$.
However, the result follows from the existence of black holes. Suppose
we start with a state whose entropy is bigger than this bound, but
whose energy is smaller than the mass of a black hole that fills the
entire box. Then we slowly add matter, increasing the energy.  At
some stage, gravitational collapse takes over and a black hole is
formed, which then continues to grow until it fills up the box.
We know that the entropy of a black hole is proportional to
the area of the horizon
measured in units of Newton's constant. However, for this black hole,
the area of the horizon is the same as the area of the box $A$, i.e.\ the
right-hand side of Equation~\ref{eq:holone}. Thus, we have managed to
decrease the entropy in this process and hence have violated the
second law. 't Hooft \cite{thooft2} and Susskind \cite{susskind2}
have proposed a radical interpretation of the
holographic bound
 (Equation~\ref{eq:holone}), namely,
a $(d+1)$-dimensional theory
of gravity should be equivalent to a $d$-dimensional theory that
lives on the boundary of space, and this $d$-dimensional theory must
have one degree of freedom per Planck area. If this is true, the
Bekenstein bound naturally follows.

We have seen that supergravity in AdS$_5 \times S^5$ is dual to
a strongly coupled large-$N$ Yang-Mills theory in $3+1$ dimensions.
Is it then true that the latter has one degree of freedom per
Planck area? This is indeed true \cite{susskindwitten}. In the
AdS$_5 \times S^5$ metric (Equation~\ref{eq:tbthreea})
the boundary is at $z = 0$ and continuum Yang-Mills theory living on this
boundary is dual to supergravity in the whole space. But suppose we
impose an infrared cutoff at $z = z_0$. Then it turns out that
supergravity in the region $z > z_0$ is dual to a Yang-Mills theory with
a position space ultraviolet cutoff $\delta = z_0$. In fact,
renormalization group flows of the gauge theory become motion in the
fifth dimension.

Consider a
large patch of this boundary of coordinate size $L$ (in terms of
the ${\vec x}$ coordinates). It follows from Equation~\ref{eq:tbthreea}
that the physical area of this boundary is
\ben
A \sim {R^3 \over z_0^3}~L^3.
\label{eq:holothree}
\een
Using the expression for $R$ in Equation~\ref{eq:tbtwo} and the fact that
the 5-dimensional Newton's
 constant $G_5$ is given by
$G_5 \sim G_{10}/R^5 \sim (g^2 l_s^8)/R^5$, we see that
\ben
{A \over G_5} \sim N^2 {L^3 \over z_0^3}.
\label{eq:holofour}
\een
Now the Yang-Mills theory has $N^2$ degrees of freedom per unit cell.
This is manifest at weak coupling, but the result for the entropy of
near-extremal
3-branes
shows that this must also be true in strong
coupling. Furthermore, the cutoff of this theory is precisely $z_0$.
Thus, the right-hand side of Equation~\ref{eq:holofour} is indeed the
number of degrees of freedom in this cutoff Yang-Mills theory. This
shows that the gauge theory--supergravity duality provides a
concrete demonstration of the Bekenstein bound.

\subsection{Near-Horizon Limit of 5D Black Hole}

%
%
Another example of the holographic connection is provided by the near-horizon geometry
of the D1-D5 system.  This spacetime is AdS$_3
\times S^3 \times M^4$ \cite{maldastromads}, where $M^4$ is either
$T^4$ or $K^3$ on which the five branes are wrapped. The AdS$_3$ is
made up of time, the transverse radial coordinate $r$,
and $x^5$. The
boundary theory is once again a superconformal
field theory. 
However, this is no longer
a gauge theory but rather the low-energy theory of
the D1-D5 system discussed in the previous sections.  Once again, the
symmetries of the bulk agree with the symmetries of the conformal
field theory. Furthermore, 5D black holes obtained by putting in some
momentum now become the Banados-Teitelboim-Zanelli (BTZ)
black holes of 3-dimensional gravity.
In fact, recently the holographic connection has been used to
understand the properties of the D1-D5 system. Details of this development
are supplied elsewhere \cite{adsreview}.

\subsection{A Suggestive Model of Holography}

Unfortunately, despite these agreements between the two dual
descriptions, we do not yet understand the explicit map that
relates the gauge theory variables to the gravity variables. There is,
in fact, an earlier example of holography: $(1+1)$-dimensional string
theory. This theory has a nonperturbative definition in terms of
quantum mechanics of $N \times N$ matrices.  One starts with the
action $S=\int dt [{\dot M}^2-V(M)]$, noting that when the
matrix is large, there is a critical form for the potential in which
successive terms in the loop expansion are comparable.
The theory can be written in terms of $\rho (x,t)$, the density
of eigenvalues of the matrix \cite{sakitajevicki}.
At the critical point, the extra dimension $x$ (which is the space of
eigenvalues) becomes continuous and the collective field $\rho (x,t)$
becomes a smooth field in $1+1$ dimensions. Remarkably, in terms of
appropriate variables,
these two dimensions (which emerged in totally
different ways) exhibit a local Lorentz invariance
\cite{dasjevicki}, though in the
presence of a background that breaks the full symmetry. In fact,
at the critical point the loop diagrams of the matrix model generate
discretizations of an oriented world sheet with dynamical world sheet
metric
and one scalar field. The usual quantization of noncritical
string theory then converts the scale on the world sheet to an extra
dimension in target space \cite{dnw},
resulting in
a $(1+1)$-dimensional
field theory. Furthermore, renormalization group flow on the world sheet
is related to motion in this extra target space direction \cite{ddw}.

It would be interesting if the gauge theory in $3+1$ dimensions
acquired extra dimensions in a similar manner to map onto string
theory in 10 dimensions.
So far, such an approach has not
succeeded, despite several attempts.

\section{DISCUSSION}

The progress of string theory over the past few years has
been truly remarkable. Consistent quantization of the string forced us
to consider a 10-dimensional supersymmetric theory. Within such a theory, one
discovers branes as solitonic states. Regarding such solitons as elementary
quanta turns out to be a duality transformation,
and such duality maps unify
all the five
perturbative  quantizations of the string, as well as 11-dimensional supergravity. There
is no room for any free parameters in the theory, although many
solutions of the field equations appear equally
acceptable at first sight,
and we do not yet know which to choose to describe the low-energy physics of
the world that we see around us.

The results on black holes have provided a striking validation of
this entire structure.
The entropy of black holes can be deduced from robust thermodynamic
arguments, using
   just the classical properties of the black hole metric and Gedanken
experiments with low-density quantum matter.
A correct quantum
   theory of gravity should reproduce this entropy by a count of allowed
microstates, and string theory yields the correct microscopic entropy
of extremal
and near-extremal holes (the only cases in which we know how to compute with
some confidence). Further, interaction with these microscopic degrees
of freedom
   yields a unitary description of black hole absorption and Hawking
emission for
low-energy quanta from near-extremal holes (again this is the case
in which we may compute with some confidence),
   thus suggesting that black holes can be
unified into physics without having to modify basic
principles of quantum mechanics or statistical physics.
In no other route
that has been attempted for quantizing gravity has it been possible
to arrive at a value for
   black hole entropy; the power of supersymmetry and our
understanding of the nonperturbative features of string theory
have allowed us to
extract more than just the perturbative physics of graviton scattering.

What about the information issue in black hole evaporation? The derivation of
   unitary low-energy Hawking radiation from microstates certainly
gives a strong indication
   on the information {\it question,} i.e.\ that information in the matter
falling into the hole is not lost but is  recovered in the  radiation
that emerges. But it is fair
   to say that we do not yet have a good understanding of the
information {\it paradox.} The paradox presents us with one way
of computing Hawking
   radiation, via the foliation of spacetime depicted in Figure~1, and this
radiation process appears to be nonunitary. The paradox then
challenges us to point out which
   of the assumptions that went into the calculations were wrong.
It  now appears reasonable that there are nonlocalities in quantum
gravity that can extend
beyond Planck scale. Indeed holography, and in particular the
Maldacena conjecture,  requires that the degrees of freedom in a
volume be encoded by
degrees of freedom at the surface of that volume, so we must give up
naive ways of thinking about locality. But we do not yet understand
in detail the mechanism
   of this nonlocality, and in particular we lack an explicit
description in string theory of the foliation in Figure~1.

These gaps in our understanding are probably linked to our lack of an
   adequate language to understand the variables of string theory in a
way that is not based
on perturbation around a given background. In recent years Matrix
theory \cite{bfss}
has attempted such a formulation, but our
understanding is still somewhat
   primitive. It thus appears that most of the progress in
understanding quantum gravity and the structure of spacetime is
yet to come, and
   string theory should be an exciting field in the years ahead.

\end{document}